\ifdefined\PaperSixThesisMode
\else
\documentclass[longauth]{aa}

\usepackage{graphicx}
\usepackage{dblfloatfix}
\usepackage{natbib}
\usepackage{scalerel}

\usepackage[table]{xcolor}
\DeclareRobustCommand{\sampleprovisional}[1]{#1}

\usepackage{siunitx}
\usepackage[percent]{overpic}
\newcommand{\numberedimage}[4][]{%
\begin{overpic}[width=#3\linewidth,#1]{#4}%
\put(4,86){\colorbox{black}{\textcolor{white}{\scriptsize\bfseries #2}}}%
\end{overpic}%
}
\newcommand{\numberedscalebarimage}[6][]{%
\begin{overpic}[width=#3\linewidth,#1]{#4}%
\put(4,86){\colorbox{black}{\textcolor{white}{\scriptsize\bfseries #2}}}%
\put(8,10){\color{white}\linethickness{1.4pt}\line(1,0){#5}}%
\put(8,14){\textcolor{white}{\scriptsize \ang{;;#6}}}%
\ifdim #3pt<0.2pt \def\compassfont{\tiny}\def\compasslines{\thinlines}\else \def\compassfont{\scriptsize}\def\compasslines{\thicklines}\fi%
{\color{white}\compasslines\put(90,8){\vector(0,1){16}}\put(90,8){\vector(-1,0){16}}}%
\put(90,27){\makebox(0,0)[b]{\textcolor{white}{\compassfont N}}}%
\put(71,8){\makebox(0,0)[r]{\textcolor{white}{\compassfont E}}}%
\end{overpic}%
}
\newcommand{\labeledmodelimage}[3]{%
\begin{overpic}[width=#2\linewidth]{#3}%
\put(3,48.5){\colorbox{white}{\textcolor{black}{\tiny\bfseries #1}}}%
\end{overpic}%
}
\newcommand{\labeledsixpanelmodelimage}[3]{%
\begin{overpic}[width=#2\linewidth]{#3}%
\put(1.2,19.5){\colorbox{white}{\textcolor{black}{\tiny\bfseries #1}}}%
\end{overpic}%
}
\newcommand{\labeledmodelimagesingleplane}[3]{%
\begin{overpic}[width=#2\linewidth]{#3}%
\put(3,22.0){\colorbox{white}{\textcolor{black}{\tiny\bfseries #1}}}%
\end{overpic}%
}

\usepackage{txfonts}
\usepackage[pdfencoding=auto,psdextra]{hyperref}
\hypersetup{
    colorlinks=true,
    linkcolor=blue,
    filecolor=magenta,      
    urlcolor=blue,
    citecolor=blue
}
\makeatletter
\renewcommand*\aa@pageof{, page \thepage{} of \pageref*{LastPage}}
\makeatother

\usepackage[utf8]{inputenc}

\usepackage[switch, modulo]{lineno}
              
\nolinenumbers

\renewcommand{\linenumbers}[0]{}

\usepackage{euclid}

\begin{document}
%
%

\title{Euclid Data Release 1 (DR1)}
\subtitle{Double-source-plane lens candidates}


\makeatletter\renewcommand*\aa@maxfirstpageauthors{21}\makeatother
\clearpage
\newcommand{\orcid}[1]{} 
\author{Euclid Collaboration: T.~Li\orcid{0009-0005-5008-0381}\thanks{\email{tian.li@port.ac.uk}}\inst{\ref{aff1}}
\and N.~E.~P.~Lines\orcid{0009-0004-7751-1914}\inst{\ref{aff1}}
\and P.~Holloway\orcid{0009-0002-8896-6100}\inst{\ref{aff1}}
\and T.~E.~Collett\orcid{0000-0001-5564-3140}\inst{\ref{aff1}}
\and J.~W.~Nightingale\orcid{0000-0002-8987-7401}\inst{\ref{aff2}}
\and S.~H.~Vincken\orcid{0009-0005-7305-2359}\inst{\ref{aff1},\ref{aff3},\ref{aff4}}
\and D.~J.~Bowden\orcid{0009-0008-6114-1401}\inst{\ref{aff5}}
\and S.~Schuldt\orcid{0000-0003-2497-6334}\inst{\ref{aff6},\ref{aff7},\ref{aff8}}
\and M.~Fogliardi\orcid{0009-0006-4964-5311}\inst{\ref{aff9},\ref{aff10}}
\and M.~Grespan\orcid{0000-0002-8213-0217}\inst{\ref{aff5},\ref{aff11}}
\and A.~Verma\orcid{0000-0002-0730-0781}\inst{\ref{aff5}}
\and C.~Krawczyk\orcid{0000-0001-9233-2341}\inst{\ref{aff1}}
\and L.~R.~Ecker\orcid{0009-0005-3508-2469}\inst{\ref{aff12},\ref{aff13},\ref{aff14}}
\and W.~J.~R.~Enzi\orcid{0009-0004-2992-3148}\inst{\ref{aff1},\ref{aff15}}
\and G.~Despali\orcid{0000-0001-6150-4112}\inst{\ref{aff16},\ref{aff10},\ref{aff17}}
\and A.~Galan\orcid{0000-0003-2547-9815}\inst{\ref{aff18}}
\and A.~Sonnenfeld\orcid{0000-0002-6061-5977}\inst{\ref{aff19},\ref{aff20},\ref{aff21}}
\and F.~Courbin\orcid{0000-0003-0758-6510}\inst{\ref{aff22},\ref{aff23},\ref{aff24}}
\and R.~B.~Metcalf\orcid{0000-0003-3167-2574}\inst{\ref{aff16},\ref{aff10}}
\and G.~Mahler\orcid{0000-0003-3266-2001}\inst{\ref{aff25}}
\and R.~J.~Massey\orcid{0000-0002-6085-3780}\inst{\ref{aff26}}
\and B.~Altieri\orcid{0000-0003-3936-0284}\inst{\ref{aff27}}
\and S.~Andreon\orcid{0000-0002-2041-8784}\inst{\ref{aff28}}
\and N.~Auricchio\orcid{0000-0003-4444-8651}\inst{\ref{aff10}}
\and C.~Baccigalupi\orcid{0000-0002-8211-1630}\inst{\ref{aff29},\ref{aff30},\ref{aff31},\ref{aff32}}
\and M.~Baldi\orcid{0000-0003-4145-1943}\inst{\ref{aff33},\ref{aff10},\ref{aff17}}
\and S.~Bardelli\orcid{0000-0002-8900-0298}\inst{\ref{aff10}}
\and P.~Battaglia\orcid{0000-0002-7337-5909}\inst{\ref{aff10}}
\and R.~Bender\orcid{0000-0001-7179-0626}\inst{\ref{aff13},\ref{aff12}}
\and A.~Biviano\orcid{0000-0002-0857-0732}\inst{\ref{aff30},\ref{aff29}}
\and E.~Branchini\orcid{0000-0002-0808-6908}\inst{\ref{aff34},\ref{aff35},\ref{aff28}}
\and M.~Brescia\orcid{0000-0001-9506-5680}\inst{\ref{aff36},\ref{aff37}}
\and J.~Brinchmann\orcid{0000-0003-4359-8797}\inst{\ref{aff38},\ref{aff39},\ref{aff40}}
\and S.~Camera\orcid{0000-0003-3399-3574}\inst{\ref{aff41},\ref{aff42},\ref{aff43}}
\and G.~Ca\~nas-Herrera\orcid{0000-0003-2796-2149}\inst{\ref{aff44}}
\and V.~Capobianco\orcid{0000-0002-3309-7692}\inst{\ref{aff43}}
\and C.~Carbone\orcid{0000-0003-0125-3563}\inst{\ref{aff8}}
\and J.~Carretero\orcid{0000-0002-3130-0204}\inst{\ref{aff45},\ref{aff46}}
\and M.~Castellano\orcid{0000-0001-9875-8263}\inst{\ref{aff47}}
\and G.~Castignani\orcid{0000-0001-6831-0687}\inst{\ref{aff10}}
\and S.~Cavuoti\orcid{0000-0002-3787-4196}\inst{\ref{aff37},\ref{aff48}}
\and A.~Cimatti\inst{\ref{aff49}}
\and C.~Colodro-Conde\inst{\ref{aff50}}
\and G.~Congedo\orcid{0000-0003-2508-0046}\inst{\ref{aff51}}
\and C.~J.~Conselice\orcid{0000-0003-1949-7638}\inst{\ref{aff52}}
\and L.~Conversi\orcid{0000-0002-6710-8476}\inst{\ref{aff53},\ref{aff27}}
\and Y.~Copin\orcid{0000-0002-5317-7518}\inst{\ref{aff54}}
\and H.~M.~Courtois\orcid{0000-0003-0509-1776}\inst{\ref{aff55}}
\and M.~Cropper\orcid{0000-0003-4571-9468}\inst{\ref{aff56}}
\and A.~Da~Silva\orcid{0000-0002-6385-1609}\inst{\ref{aff57},\ref{aff58}}
\and H.~Degaudenzi\orcid{0000-0002-5887-6799}\inst{\ref{aff18}}
\and G.~De~Lucia\orcid{0000-0002-6220-9104}\inst{\ref{aff30}}
\and C.~Dolding\orcid{0009-0003-7199-6108}\inst{\ref{aff56}}
\and H.~Dole\orcid{0000-0002-9767-3839}\inst{\ref{aff59},\ref{aff60}}
\and F.~Dubath\orcid{0000-0002-6533-2810}\inst{\ref{aff18}}
\and X.~Dupac\inst{\ref{aff27}}
\and S.~Escoffier\orcid{0000-0002-2847-7498}\inst{\ref{aff61}}
\and M.~Fabricius\orcid{0000-0002-7025-6058}\inst{\ref{aff13},\ref{aff12}}
\and M.~Farina\orcid{0000-0002-3089-7846}\inst{\ref{aff62}}
\and R.~Farinelli\inst{\ref{aff10}}
\and F.~Faustini\orcid{0000-0001-6274-5145}\inst{\ref{aff47}}
\and S.~Ferriol\inst{\ref{aff54}}
\and M.~Frailis\orcid{0000-0002-7400-2135}\inst{\ref{aff30}}
\and E.~Franceschi\orcid{0000-0002-0585-6591}\inst{\ref{aff10}}
\and M.~Fumana\orcid{0000-0001-6787-5950}\inst{\ref{aff8}}
\and L.~Gabarra\orcid{0000-0002-8486-8856}\inst{\ref{aff5}}
\and S.~Galeotta\orcid{0000-0002-3748-5115}\inst{\ref{aff30}}
\and K.~George\orcid{0000-0002-1734-8455}\inst{\ref{aff63}}
\and C.~Giocoli\orcid{0000-0002-9590-7961}\inst{\ref{aff10},\ref{aff17}}
\and J.~Gracia-Carpio\orcid{0000-0003-4689-3134}\inst{\ref{aff13}}
\and A.~Grazian\orcid{0000-0002-5688-0663}\inst{\ref{aff64}}
\and F.~Grupp\inst{\ref{aff13},\ref{aff12}}
\and L.~Guzzo\orcid{0000-0001-8264-5192}\inst{\ref{aff65},\ref{aff28},\ref{aff66}}
\and W.~G.~Hartley\orcid{0000-0002-4563-5101}\inst{\ref{aff18}}
\and S.~V.~H.~Haugan\orcid{0000-0001-9648-7260}\inst{\ref{aff67}}
\and S.~Hemmati\orcid{0000-0003-2226-5395}\inst{\ref{aff68}}
\and H.~Hoekstra\orcid{0000-0002-0641-3231}\inst{\ref{aff44}}
\and W.~Holmes\orcid{0009-0007-8554-4646}\inst{\ref{aff69}}
\and I.~M.~Hook\orcid{0000-0002-2960-978X}\inst{\ref{aff70}}
\and F.~Hormuth\inst{\ref{aff71}}
\and A.~Hornstrup\orcid{0000-0002-3363-0936}\inst{\ref{aff72},\ref{aff73}}
\and K.~Jahnke\orcid{0000-0003-3804-2137}\inst{\ref{aff74}}
\and M.~Jhabvala\inst{\ref{aff75}}
\and A.~Kiessling\orcid{0000-0002-2590-1273}\inst{\ref{aff69}}
\and B.~Kubik\orcid{0009-0006-5823-4880}\inst{\ref{aff54}}
\and M.~K\"ummel\orcid{0000-0003-2791-2117}\inst{\ref{aff12}}
\and M.~Kunz\orcid{0000-0002-3052-7394}\inst{\ref{aff76}}
\and H.~Kurki-Suonio\orcid{0000-0002-4618-3063}\inst{\ref{aff7},\ref{aff77}}
\and A.~M.~C.~Le~Brun\orcid{0000-0002-0936-4594}\inst{\ref{aff78}}
\and P.~Liebing\orcid{0009-0005-8043-7150}\inst{\ref{aff56}}
\and S.~Ligori\orcid{0000-0003-4172-4606}\inst{\ref{aff43}}
\and P.~B.~Lilje\orcid{0000-0003-4324-7794}\inst{\ref{aff67}}
\and V.~Lindholm\orcid{0000-0003-2317-5471}\inst{\ref{aff7},\ref{aff77}}
\and I.~Lloro\orcid{0000-0001-5966-1434}\inst{\ref{aff79}}
\and M.~Magliocchetti\orcid{0000-0001-9158-4838}\inst{\ref{aff62}}
\and G.~Mainetti\orcid{0000-0003-2384-2377}\inst{\ref{aff80}}
\and O.~Marggraf\orcid{0000-0001-7242-3852}\inst{\ref{aff81}}
\and M.~Martinelli\orcid{0000-0002-6943-7732}\inst{\ref{aff47},\ref{aff82}}
\and N.~Martinet\orcid{0000-0003-2786-7790}\inst{\ref{aff83}}
\and F.~Marulli\orcid{0000-0002-8850-0303}\inst{\ref{aff16},\ref{aff10},\ref{aff17}}
\and N.~Mauri\orcid{0000-0001-8196-1548}\inst{\ref{aff49},\ref{aff17}}
\and E.~Medinaceli\orcid{0000-0002-4040-7783}\inst{\ref{aff10}}
\and M.~Melchior\inst{\ref{aff4}}
\and M.~Meneghetti\orcid{0000-0003-1225-7084}\inst{\ref{aff10},\ref{aff17}}
\and G.~Meylan\orcid{0000-0001-6503-0209}\inst{\ref{aff84}}
\and P.~Monaco\orcid{0000-0003-2083-7564}\inst{\ref{aff85},\ref{aff30},\ref{aff31},\ref{aff29}}
\and M.~Moresco\orcid{0000-0002-7616-7136}\inst{\ref{aff16},\ref{aff10}}
\and C.~Moretti\orcid{0000-0003-3314-8936}\inst{\ref{aff30},\ref{aff29},\ref{aff31}}
\and L.~Moscardini\orcid{0000-0002-3473-6716}\inst{\ref{aff16},\ref{aff10},\ref{aff17}}
\and R.~Nakajima\orcid{0009-0009-1213-7040}\inst{\ref{aff81}}
\and C.~Neissner\orcid{0000-0001-8524-4968}\inst{\ref{aff86},\ref{aff46}}
\and R.~C.~Nichol\orcid{0000-0003-0939-6518}\inst{\ref{aff87}}
\and S.-M.~Niemi\orcid{0009-0005-0247-0086}\inst{\ref{aff88}}
\and C.~Padilla\orcid{0000-0001-7951-0166}\inst{\ref{aff86}}
\and S.~Paltani\orcid{0000-0002-8108-9179}\inst{\ref{aff18}}
\and F.~Pasian\orcid{0000-0002-4869-3227}\inst{\ref{aff30}}
\and W.~J.~Percival\orcid{0000-0002-0644-5727}\inst{\ref{aff89},\ref{aff90},\ref{aff91}}
\and V.~Pettorino\orcid{0000-0002-4203-9320}\inst{\ref{aff88}}
\and S.~Pires\orcid{0000-0002-0249-2104}\inst{\ref{aff92}}
\and G.~Polenta\orcid{0000-0003-4067-9196}\inst{\ref{aff93}}
\and M.~Poncet\inst{\ref{aff94}}
\and G.~D.~Racca\orcid{0000-0002-9883-8981}\inst{\ref{aff44},\ref{aff88}}
\and F.~Raison\orcid{0000-0002-7819-6918}\inst{\ref{aff13}}
\and A.~Renzi\orcid{0000-0001-9856-1970}\inst{\ref{aff95},\ref{aff96},\ref{aff10}}
\and J.~Rhodes\orcid{0000-0002-4485-8549}\inst{\ref{aff69}}
\and G.~Riccio\inst{\ref{aff37}}
\and I.~Risso\orcid{0000-0003-2525-7761}\inst{\ref{aff34},\ref{aff35},\ref{aff28}}
\and F.~Rizzo\orcid{0000-0002-9407-585X}\inst{\ref{aff30}}
\and E.~Romelli\orcid{0000-0003-3069-9222}\inst{\ref{aff30}}
\and M.~Roncarelli\orcid{0000-0001-9587-7822}\inst{\ref{aff10}}
\and B.~Rusholme\orcid{0000-0001-7648-4142}\inst{\ref{aff68}}
\and R.~Saglia\orcid{0000-0003-0378-7032}\inst{\ref{aff12},\ref{aff13}}
\and Z.~Sakr\orcid{0000-0002-4823-3757}\inst{\ref{aff97},\ref{aff98},\ref{aff99}}
\and D.~Sapone\orcid{0000-0001-7089-4503}\inst{\ref{aff100}}
\and P.~Schneider\orcid{0000-0001-8561-2679}\inst{\ref{aff81}}
\and A.~Secroun\orcid{0000-0003-0505-3710}\inst{\ref{aff61}}
\and E.~Sihvola\orcid{0000-0003-1804-7715}\inst{\ref{aff7},\ref{aff77}}
\and P.~Simon\inst{\ref{aff81}}
\and C.~Sirignano\orcid{0000-0002-0995-7146}\inst{\ref{aff95},\ref{aff96}}
\and G.~Sirri\orcid{0000-0003-2626-2853}\inst{\ref{aff17}}
\and L.~Stanco\orcid{0000-0002-9706-5104}\inst{\ref{aff96}}
\and P.~Tallada-Cresp\'{i}\orcid{0000-0002-1336-8328}\inst{\ref{aff45},\ref{aff46}}
\and A.~N.~Taylor\inst{\ref{aff51}}
\and H.~I.~Teplitz\orcid{0000-0002-7064-5424}\inst{\ref{aff101}}
\and I.~Tereno\orcid{0000-0002-4537-6218}\inst{\ref{aff57},\ref{aff58}}
\and S.~Toft\orcid{0000-0003-3631-7176}\inst{\ref{aff102},\ref{aff103}}
\and R.~Toledo-Moreo\orcid{0000-0002-2997-4859}\inst{\ref{aff104},\ref{aff105}}
\and F.~Torradeflot\orcid{0000-0003-1160-1517}\inst{\ref{aff46},\ref{aff45}}
\and I.~Tutusaus\orcid{0000-0002-3199-0399}\inst{\ref{aff106},\ref{aff107},\ref{aff98}}
\and J.~Valiviita\orcid{0000-0001-6225-3693}\inst{\ref{aff7},\ref{aff77}}
\and T.~Vassallo\orcid{0000-0001-6512-6358}\inst{\ref{aff30},\ref{aff63}}
\and A.~Veropalumbo\orcid{0000-0003-2387-1194}\inst{\ref{aff28},\ref{aff35},\ref{aff34}}
\and Y.~Wang\orcid{0000-0002-4749-2984}\inst{\ref{aff68}}
\and J.~Weller\orcid{0000-0002-8282-2010}\inst{\ref{aff12},\ref{aff13}}
\and A.~Zacchei\orcid{0000-0003-0396-1192}\inst{\ref{aff30},\ref{aff29}}
\and G.~Zamorani\orcid{0000-0002-2318-301X}\inst{\ref{aff10}}
\and F.~M.~Zerbi\orcid{0000-0002-9996-973X}\inst{\ref{aff28}}
\and E.~Zucca\orcid{0000-0002-5845-8132}\inst{\ref{aff10}}
\and I.~T.~Andika\orcid{0000-0001-6102-9526}\inst{\ref{aff12}}
\and M.~Ballardini\orcid{0000-0003-4481-3559}\inst{\ref{aff9},\ref{aff108},\ref{aff10}}
\and P.~Bergamini\orcid{0000-0003-1383-9414}\inst{\ref{aff10}}
\and S.~Borgani\orcid{0000-0001-6151-6439}\inst{\ref{aff85},\ref{aff29},\ref{aff30},\ref{aff31},\ref{aff109}}
\and E.~Bozzo\orcid{0000-0002-8201-1525}\inst{\ref{aff18}}
\and C.~Burigana\orcid{0000-0002-3005-5796}\inst{\ref{aff15},\ref{aff110}}
\and M.~Calabrese\orcid{0000-0002-2637-2422}\inst{\ref{aff111},\ref{aff8}}
\and F.~Caro\orcid{0009-0003-1053-0507}\inst{\ref{aff47}}
\and D.~Di~Ferdinando\inst{\ref{aff17}}
\and J.~A.~Escartin~Vigo\inst{\ref{aff13}}
\and J.~Garc\'ia-Bellido\orcid{0000-0002-9370-8360}\inst{\ref{aff97}}
\and T.~Gasparetto\orcid{0000-0002-7913-4866}\inst{\ref{aff47}}
\and C.~M.~Gutierrez\orcid{0000-0001-7854-783X}\inst{\ref{aff50},\ref{aff112}}
\and A.~Hall\orcid{0000-0002-3139-8651}\inst{\ref{aff51}}
\and Y.~Kang\orcid{0009-0000-8588-7250}\inst{\ref{aff113}}
\and A.~Loureiro\orcid{0000-0002-4371-0876}\inst{\ref{aff114},\ref{aff115}}
\and R.~Maoli\orcid{0000-0002-6065-3025}\inst{\ref{aff116},\ref{aff47}}
\and S.~Nadathur\orcid{0000-0001-9070-3102}\inst{\ref{aff1}}
\and M.~P\"ontinen\orcid{0000-0001-5442-2530}\inst{\ref{aff7}}
\and E.~Sarpa\orcid{0000-0002-1256-655X}\inst{\ref{aff30}}
\and V.~Scottez\orcid{0009-0008-3864-940X}\inst{\ref{aff117},\ref{aff118}}
\and M.~Sereno\orcid{0000-0003-0302-0325}\inst{\ref{aff10},\ref{aff17}}
\and K.~Tanidis\orcid{0000-0001-9843-5130}\inst{\ref{aff119}}
\and M.~Tenti\orcid{0000-0002-4254-5901}\inst{\ref{aff17}}
\and A.~Troja\orcid{0000-0003-0239-4595}\inst{\ref{aff30}}
\and M.~Tucci\inst{\ref{aff18}}
\and M.~Viel\orcid{0000-0002-2642-5707}\inst{\ref{aff29},\ref{aff30},\ref{aff32},\ref{aff31},\ref{aff109}}
\and M.~Wiesmann\orcid{0009-0000-8199-5860}\inst{\ref{aff67}}
\and Abdurro'uf\orcid{0000-0002-5258-8761}\inst{\ref{aff120}}
\and J.~A.~Acevedo~Barroso\orcid{0000-0002-9654-1711}\inst{\ref{aff69}}
\and Y.~Akrami\orcid{0000-0002-2407-7956}\inst{\ref{aff97},\ref{aff121}}
\and G.~Angora\orcid{0000-0002-0316-6562}\inst{\ref{aff37},\ref{aff9}}
\and M.~Archidiacono\orcid{0000-0003-4952-9012}\inst{\ref{aff65},\ref{aff66}}
\and G.~Aric\`o\orcid{0000-0002-2802-2928}\inst{\ref{aff17}}
\and F.~Atrio-Barandela\orcid{0000-0002-2130-2513}\inst{\ref{aff122}}
\and M.~Aubert\orcid{0009-0002-7667-8814}\inst{\ref{aff54}}
\and M.~Baes\orcid{0000-0002-3930-2757}\inst{\ref{aff123}}
\and N.~Bartolo\inst{\ref{aff95},\ref{aff96},\ref{aff64}}
\and L.~Bazzanini\orcid{0000-0003-0727-0137}\inst{\ref{aff15},\ref{aff9}}
\and J.~R.~Bermejo-Climent\inst{\ref{aff124},\ref{aff125}}
\and D.~Bertacca\orcid{0000-0002-2490-7139}\inst{\ref{aff95},\ref{aff64},\ref{aff96}}
\and M.~Bethermin\orcid{0000-0002-3915-2015}\inst{\ref{aff126}}
\and F.~Beutler\orcid{0000-0003-0467-5438}\inst{\ref{aff51}}
\and A.~Blanchard\orcid{0000-0001-8555-9003}\inst{\ref{aff98}}
\and F.~Bouch\`e\orcid{0000-0002-4663-1786}\inst{\ref{aff127},\ref{aff48}}
\and M.~L.~Brown\orcid{0000-0002-0370-8077}\inst{\ref{aff52}}
\and S.~Bruton\orcid{0000-0002-6503-5218}\inst{\ref{aff128}}
\and A.~Calabro\orcid{0000-0003-2536-1614}\inst{\ref{aff47}}
\and C.~S.~Carvalho\inst{\ref{aff129}}
\and N.~Chartab\orcid{0000-0003-3691-937X}\inst{\ref{aff68}}
\and F.~Cogato\orcid{0000-0003-4632-6113}\inst{\ref{aff10}}
\and P.~Corcho-Caballero\orcid{0000-0001-6327-7080}\inst{\ref{aff130}}
\and M.~Correnti\orcid{0000-0001-6464-3257}\inst{\ref{aff47},\ref{aff93}}
\and B.~Csizi\orcid{0000-0003-3227-6581}\inst{\ref{aff113}}
\and O.~Cucciati\orcid{0000-0002-9336-7551}\inst{\ref{aff10}}
\and T.~de~Boer\orcid{0000-0001-5486-2747}\inst{\ref{aff131}}
\and F.~De~Paolis\orcid{0000-0001-6460-7563}\inst{\ref{aff132},\ref{aff133},\ref{aff134}}
\and G.~Desprez\orcid{0000-0001-8325-1742}\inst{\ref{aff130}}
\and A.~D\'iaz-S\'anchez\orcid{0000-0003-0748-4768}\inst{\ref{aff135}}
\and S.~Di~Domizio\orcid{0000-0003-2863-5895}\inst{\ref{aff34},\ref{aff35}}
\and J.~M.~Diego\orcid{0000-0001-9065-3926}\inst{\ref{aff136}}
\and P.-A.~Duc\orcid{0000-0003-3343-6284}\inst{\ref{aff126}}
\and V.~Duret\orcid{0009-0009-0383-4960}\inst{\ref{aff137}}
\and Y.~Fang\orcid{0000-0002-0334-6950}\inst{\ref{aff12}}
\and A.~Farina\orcid{0009-0000-3420-929X}\inst{\ref{aff28},\ref{aff35}}
\and A.~Finoguenov\orcid{0000-0002-4606-5403}\inst{\ref{aff7}}
\and A.~Fontana\orcid{0000-0003-3820-2823}\inst{\ref{aff47}}
\and A.~Franco\orcid{0000-0002-4761-366X}\inst{\ref{aff132},\ref{aff133},\ref{aff134}}
\and N.~Frusciante\orcid{0000-0002-7375-1230}\inst{\ref{aff138},\ref{aff139}}
\and Y.~Fu\orcid{0000-0002-0759-0504}\inst{\ref{aff44},\ref{aff130}}
\and K.~Ganga\orcid{0000-0001-8159-8208}\inst{\ref{aff140}}
\and R.~Gavazzi\orcid{0000-0002-5540-6935}\inst{\ref{aff83},\ref{aff141}}
\and E.~Gaztanaga\orcid{0000-0001-9632-0815}\inst{\ref{aff106},\ref{aff107},\ref{aff1}}
\and Z.~Ghaffari\orcid{0000-0002-6467-8078}\inst{\ref{aff30},\ref{aff29}}
\and F.~Giacomini\orcid{0000-0002-3129-2814}\inst{\ref{aff17}}
\and F.~Gianotti\orcid{0000-0003-4666-119X}\inst{\ref{aff10}}
\and G.~Gozaliasl\orcid{0000-0002-0236-919X}\inst{\ref{aff142},\ref{aff7}}
\and A.~Gruppuso\orcid{0000-0001-9272-5292}\inst{\ref{aff10},\ref{aff17}}
\and M.~Guidi\orcid{0000-0001-9408-1101}\inst{\ref{aff33},\ref{aff10}}
\and C.~Hern\'andez-Monteagudo\orcid{0000-0001-5471-9166}\inst{\ref{aff112},\ref{aff50}}
\and T.~Heuberger\orcid{0009-0009-3474-8467}\inst{\ref{aff143}}
\and J.~Hjorth\orcid{0000-0002-4571-2306}\inst{\ref{aff144}}
\and M.~Jauzac\orcid{0000-0003-1974-8732}\inst{\ref{aff145},\ref{aff26},\ref{aff146},\ref{aff147},\ref{aff98}}
\and S.~Joudaki\orcid{0000-0001-8820-673X}\inst{\ref{aff45}}
\and J.~J.~E.~Kajava\orcid{0000-0002-3010-8333}\inst{\ref{aff148},\ref{aff6},\ref{aff149}}
\and D.~Karagiannis\orcid{0000-0002-4927-0816}\inst{\ref{aff150},\ref{aff9},\ref{aff151}}
\and J.~Kim\orcid{0000-0003-2776-2761}\inst{\ref{aff5}}
\and C.~C.~Kirkpatrick\inst{\ref{aff152}}
\and A.~Kov\'acs\orcid{0000-0002-5825-579X}\inst{\ref{aff125},\ref{aff124}}
\and I.~Kova{\v{c}}i{\'{c}}\orcid{0000-0001-6751-3263}\inst{\ref{aff123}}
\and K.~Koyama\orcid{0000-0001-6727-6915}\inst{\ref{aff1}}
\and S.~Kruk\orcid{0000-0001-8010-8879}\inst{\ref{aff27}}
\and M.~C.~Lam\orcid{0000-0002-9347-2298}\inst{\ref{aff51}}
\and S.~Lee\orcid{0000-0002-8289-740X}\inst{\ref{aff153},\ref{aff69}}
\and M.~D.~Lepinzan\orcid{0000-0003-1287-9801}\inst{\ref{aff30}}
\and G.~F.~Lesci\orcid{0000-0002-4607-2830}\inst{\ref{aff15}}
\and J.~Lesgourgues\orcid{0000-0001-7627-353X}\inst{\ref{aff154}}
\and T.~I.~Liaudat\orcid{0000-0002-9104-314X}\inst{\ref{aff155}}
\and L.~Linke\orcid{0000-0002-2622-8113}\inst{\ref{aff113}}
\and S.~J.~Liu\orcid{0000-0001-7680-2139}\inst{\ref{aff62}}
\and E.~A.~Magnier\orcid{0000-0002-7965-2815}\inst{\ref{aff131}}
\and A.~Manj\'on-Garc\'ia\orcid{0000-0002-7413-8825}\inst{\ref{aff135}}
\and F.~Mannucci\orcid{0000-0002-4803-2381}\inst{\ref{aff156}}
\and F.~R.~Marleau\orcid{0000-0002-1442-2947}\inst{\ref{aff113}}
\and C.~J.~A.~P.~Martins\orcid{0000-0002-4886-9261}\inst{\ref{aff157},\ref{aff39}}
\and M.~Miluzio\inst{\ref{aff27},\ref{aff158}}
\and G.~Morgante\inst{\ref{aff10}}
\and C.~Murray\orcid{0000-0002-4668-1273}\inst{\ref{aff92}}
\and K.~Naidoo\orcid{0000-0002-9182-1802}\inst{\ref{aff1},\ref{aff74}}
\and P.~Natoli\orcid{0000-0003-0126-9100}\inst{\ref{aff9},\ref{aff108}}
\and A.~Navarro-Alsina\orcid{0000-0002-3173-2592}\inst{\ref{aff81}}
\and S.~Nesseris\orcid{0000-0002-0567-0324}\inst{\ref{aff97}}
\and L.~Nicastro\orcid{0000-0001-8534-6788}\inst{\ref{aff10}}
\and M.~Oguri\orcid{0000-0003-3484-399X}\inst{\ref{aff159},\ref{aff160}}
\and F.~Oppizzi\orcid{0000-0003-3904-8370}\inst{\ref{aff35},\ref{aff96},\ref{aff95}}
\and F.~Pace\orcid{0000-0001-8039-0480}\inst{\ref{aff41},\ref{aff42},\ref{aff43},\ref{aff58}}
\and D.~Paoletti\orcid{0000-0003-4761-6147}\inst{\ref{aff10},\ref{aff110}}
\and K.~Paterson\orcid{0000-0001-8340-3486}\inst{\ref{aff74}}
\and C.~Pattison\orcid{0000-0003-3272-2617}\inst{\ref{aff1}}
\and L.~Pizzuti\orcid{0000-0001-5654-7580}\inst{\ref{aff161}}
\and D.~Potter\orcid{0000-0002-0757-5195}\inst{\ref{aff162}}
\and G.~W.~Pratt\inst{\ref{aff92}}
\and S.~Quai\orcid{0000-0002-0449-8163}\inst{\ref{aff16},\ref{aff10}}
\and M.~Radovich\orcid{0000-0002-3585-866X}\inst{\ref{aff64}}
\and F.~Ricci\orcid{0000-0001-5742-5980}\inst{\ref{aff163},\ref{aff47}}
\and K.~Rojas\orcid{0000-0003-1391-6854}\inst{\ref{aff143}}
\and W.~Roster\orcid{0000-0002-9149-6528}\inst{\ref{aff13}}
\and M.~Sahl\'en\orcid{0000-0003-0973-4804}\inst{\ref{aff164}}
\and D.~B.~Sanders\orcid{0000-0002-1233-9998}\inst{\ref{aff131}}
\and A.~Schneider\orcid{0000-0001-7055-8104}\inst{\ref{aff162}}
\and D.~Sciotti\orcid{0009-0008-4519-2620}\inst{\ref{aff47},\ref{aff82}}
\and E.~Sellentin\orcid{0009-0002-2655-3458}\inst{\ref{aff165},\ref{aff44}}
\and S.~Serjeant\orcid{0000-0002-0517-7943}\inst{\ref{aff166}}
\and F.~Shankar\orcid{0000-0001-8973-5051}\inst{\ref{aff167}}
\and L.~C.~Smith\orcid{0000-0002-3259-2771}\inst{\ref{aff168}}
\and I.~Szapudi\orcid{0000-0003-2274-0301}\inst{\ref{aff131}}
\and G.~Testera\orcid{0000-0003-2970-766X}\inst{\ref{aff35}}
\and R.~Teyssier\orcid{0000-0001-7689-0933}\inst{\ref{aff169}}
\and S.~Tosi\orcid{0000-0002-7275-9193}\inst{\ref{aff34},\ref{aff28},\ref{aff35}}
\and C.~Uhlemann\orcid{0000-0001-7831-1579}\inst{\ref{aff170}}
\and C.~Valieri\inst{\ref{aff17}}
\and A.~Venhola\orcid{0000-0001-6071-4564}\inst{\ref{aff171}}
\and D.~Vergani\orcid{0000-0003-0898-2216}\inst{\ref{aff10}}
\and G.~Vernardos\orcid{0000-0001-8554-7248}\inst{\ref{aff172},\ref{aff173}}
\and S.~Vinciguerra\orcid{0009-0005-4018-3184}\inst{\ref{aff83}}
\and M.~von~Wietersheim-Kramsta\orcid{0000-0003-4986-5091}\inst{\ref{aff26},\ref{aff145}}
\and N.~A.~Walton\orcid{0000-0003-3983-8778}\inst{\ref{aff168}}
\and A.~H.~Wright\orcid{0000-0001-7363-7932}\inst{\ref{aff174}}}
										   
\institute{Institute of Cosmology and Gravitation, University of Portsmouth, Portsmouth PO1 3FX, UK\label{aff1}
\and
School of Mathematics, Statistics and Physics, Newcastle University, Herschel Building, Newcastle-upon-Tyne, NE1 7RU, UK\label{aff2}
\and
Fontys University of Applied Sciences, School of ICT, Rachelsmolen 1, 5612 MA Eindhoven, The Netherlands\label{aff3}
\and
University of Applied Sciences and Arts of Northwestern Switzerland, School of Engineering, 5210 Windisch, Switzerland\label{aff4}
\and
Department of Physics, University of Oxford, Keble Road, Oxford OX1 3RH, UK\label{aff5}
\and
Finnish Centre for Astronomy with ESO (FINCA), Quantum, Vesilinnantie 5, University of Turku, 20014 Turku, Finland\label{aff6}
\and
Department of Physics, P.O. Box 64, University of Helsinki, 00014 Helsinki, Finland\label{aff7}
\and
INAF-IASF Milano, Via Alfonso Corti 12, 20133 Milano, Italy\label{aff8}
\and
Dipartimento di Fisica e Scienze della Terra, Universit\`a degli Studi di Ferrara, Via Giuseppe Saragat 1, 44122 Ferrara, Italy\label{aff9}
\and
INAF-Osservatorio di Astrofisica e Scienza dello Spazio di Bologna, Via Piero Gobetti 93/3, 40129 Bologna, Italy\label{aff10}
\and
National Centre for Nuclear Research, ul. Pasteura 7, 02-093, Warsaw, Poland\label{aff11}
\and
Universit\"ats-Sternwarte M\"unchen, Fakult\"at f\"ur Physik, Ludwig-Maximilians-Universit\"at M\"unchen, Scheinerstr.~1, 81679 M\"unchen, Germany\label{aff12}
\and
Max Planck Institute for Extraterrestrial Physics, Giessenbachstr. 1, 85748 Garching, Germany\label{aff13}
\and
Max-Planck-Institut f\"ur Astrophysik, Karl-Schwarzschild-Str.~1, 85748 Garching, Germany\label{aff14}
\and
INAF, Istituto di Radioastronomia, Via Piero Gobetti 101, 40129 Bologna, Italy\label{aff15}
\and
Dipartimento di Fisica e Astronomia "Augusto Righi" - Alma Mater Studiorum Universit\`a di Bologna, via Piero Gobetti 93/2, 40129 Bologna, Italy\label{aff16}
\and
INFN-Sezione di Bologna, Viale Berti Pichat 6/2, 40127 Bologna, Italy\label{aff17}
\and
Department of Astronomy, University of Geneva, ch. d'Ecogia 16, 1290 Versoix, Switzerland\label{aff18}
\and
Department of Astronomy, School of Physics and Astronomy, Shanghai Jiao Tong University, Shanghai 200240, China\label{aff19}
\and
State Key Laboratory of Dark Matter Physics, School of Physics and Astronomy, Shanghai Jiao Tong University, Shanghai 200240, China\label{aff20}
\and
Shanghai Key Laboratory for Particle Physics and Cosmology, Shanghai 200240, China\label{aff21}
\and
Institut de Ci\`{e}ncies del Cosmos (ICCUB), Universitat de Barcelona (IEEC-UB), Mart\'{i} i Franqu\`{e}s 1, 08028 Barcelona, Spain\label{aff22}
\and
Instituci\'o Catalana de Recerca i Estudis Avan\c{c}ats (ICREA), Passeig de Llu\'{\i}s Companys 23, 08010 Barcelona, Spain\label{aff23}
\and
Institut de Ciencies de l'Espai (IEEC-CSIC), Campus UAB, Carrer de Can Magrans, s/n Cerdanyola del Vall\'es, 08193 Barcelona, Spain\label{aff24}
\and
STAR Institute, University of Li{\`e}ge, Quartier Agora, All\'ee du six Ao\^ut 19c, 4000 Li\`ege, Belgium\label{aff25}
\and
Department of Physics, Institute for Computational Cosmology, Durham University, South Road, Durham, DH1 3LE, UK\label{aff26}
\and
ESAC/ESA, Camino Bajo del Castillo, s/n., Urb. Villafranca del Castillo, 28692 Villanueva de la Ca\~nada, Madrid, Spain\label{aff27}
\and
INAF-Osservatorio Astronomico di Brera, Via Brera 28, 20122 Milano, Italy\label{aff28}
\and
IFPU, Institute for Fundamental Physics of the Universe, via Beirut 2, 34151 Trieste, Italy\label{aff29}
\and
INAF-Osservatorio Astronomico di Trieste, Via G. B. Tiepolo 11, 34143 Trieste, Italy\label{aff30}
\and
INFN, Sezione di Trieste, Via Valerio 2, 34127 Trieste TS, Italy\label{aff31}
\and
SISSA, International School for Advanced Studies, Via Bonomea 265, 34136 Trieste TS, Italy\label{aff32}
\and
Dipartimento di Fisica e Astronomia, Universit\`a di Bologna, Via Gobetti 93/2, 40129 Bologna, Italy\label{aff33}
\and
Dipartimento di Fisica, Universit\`a di Genova, Via Dodecaneso 33, 16146, Genova, Italy\label{aff34}
\and
INFN-Sezione di Genova, Via Dodecaneso 33, 16146, Genova, Italy\label{aff35}
\and
Department of Physics "E. Pancini", University Federico II, Via Cinthia 6, 80126, Napoli, Italy\label{aff36}
\and
INAF-Osservatorio Astronomico di Capodimonte, Via Moiariello 16, 80131 Napoli, Italy\label{aff37}
\and
European Southern Observatory, Karl-Schwarzschild-Str.~2, 85748 Garching, Germany\label{aff38}
\and
Instituto de Astrof\'isica e Ci\^encias do Espa\c{c}o, Universidade do Porto, CAUP, Rua das Estrelas, PT4150-762 Porto, Portugal\label{aff39}
\and
Faculdade de Ci\^encias da Universidade do Porto, Rua do Campo de Alegre, 4150-007 Porto, Portugal\label{aff40}
\and
Dipartimento di Fisica, Universit\`a degli Studi di Torino, Via P. Giuria 1, 10125 Torino, Italy\label{aff41}
\and
INFN-Sezione di Torino, Via P. Giuria 1, 10125 Torino, Italy\label{aff42}
\and
INAF-Osservatorio Astrofisico di Torino, Via Osservatorio 20, 10025 Pino Torinese (TO), Italy\label{aff43}
\and
Leiden Observatory, Leiden University, Einsteinweg 55, 2333 CC Leiden, The Netherlands\label{aff44}
\and
Centro de Investigaciones Energ\'eticas, Medioambientales y Tecnol\'ogicas (CIEMAT), Avenida Complutense 40, 28040 Madrid, Spain\label{aff45}
\and
Port d'Informaci\'{o} Cient\'{i}fica, Campus UAB, C. Albareda s/n, 08193 Bellaterra (Barcelona), Spain\label{aff46}
\and
INAF-Osservatorio Astronomico di Roma, Via Frascati 33, 00078 Monteporzio Catone, Italy\label{aff47}
\and
INFN -- Sezione di Napoli, Via Cinthia 6, 80126, Napoli, Italy\label{aff48}
\and
Dipartimento di Fisica e Astronomia "Augusto Righi" - Alma Mater Studiorum Universit\`a di Bologna, Viale Berti Pichat 6/2, 40127 Bologna, Italy\label{aff49}
\and
Instituto de Astrof\'{\i}sica de Canarias, E-38205 La Laguna, Tenerife, Spain\label{aff50}
\and
Institute for Astronomy, University of Edinburgh, Royal Observatory, Blackford Hill, Edinburgh EH9 3HJ, UK\label{aff51}
\and
Jodrell Bank Centre for Astrophysics, Department of Physics and Astronomy, University of Manchester, Oxford Road, Manchester M13 9PL, UK\label{aff52}
\and
European Space Agency/ESRIN, Largo Galileo Galilei 1, 00044 Frascati, Roma, Italy\label{aff53}
\and
Universit\'e Lyon 1, CNRS/IN2P3, IP2I Lyon, Villeurbanne, France\label{aff54}
\and
Universit\'e Lyon 1, CNRS/IN2P3, IUF, IP2I Lyon, Villeurbanne, France \label{aff55}
\and
Mullard Space Science Laboratory, University College London, Holmbury St Mary, Dorking, Surrey RH5 6NT, UK\label{aff56}
\and
Departamento de F\'isica, Faculdade de Ci\^encias, Universidade de Lisboa, Edif\'icio C8, Campo Grande, PT1749-016 Lisboa, Portugal\label{aff57}
\and
Instituto de Astrof\'isica e Ci\^encias do Espa\c{c}o, Faculdade de Ci\^encias, Universidade de Lisboa, Campo Grande, 1749-016 Lisboa, Portugal\label{aff58}
\and
Universit\'e Paris-Saclay, CNRS, Institut d'astrophysique spatiale, 91405, Orsay, France\label{aff59}
\and
Institut universitaire de France (IUF), 1 rue Descartes, 75231 PARIS CEDEX 05, France\label{aff60}
\and
Aix-Marseille Universit\'e, CNRS/IN2P3, CPPM, Marseille, France\label{aff61}
\and
INAF-Istituto di Astrofisica e Planetologia Spaziali, via del Fosso del Cavaliere, 100, 00100 Roma, Italy\label{aff62}
\and
University Observatory, LMU Faculty of Physics, Scheinerstr.~1, 81679 Munich, Germany\label{aff63}
\and
INAF-Osservatorio Astronomico di Padova, Via dell'Osservatorio 5, 35122 Padova, Italy\label{aff64}
\and
Dipartimento di Fisica "Aldo Pontremoli", Universit\`a degli Studi di Milano, Via Celoria 16, 20133 Milano, Italy\label{aff65}
\and
INFN-Sezione di Milano, Via Celoria 16, 20133 Milano, Italy\label{aff66}
\and
Institute of Theoretical Astrophysics, University of Oslo, P.O. Box 1029 Blindern, 0315 Oslo, Norway\label{aff67}
\and
Caltech/IPAC, 1200 E. California Blvd., Pasadena, CA 91125, USA\label{aff68}
\and
Jet Propulsion Laboratory, California Institute of Technology, 4800 Oak Grove Drive, Pasadena, CA, 91109, USA\label{aff69}
\and
School of Physics and Astronomy, Lancaster University, Lancaster, LA1 4YB, UK\label{aff70}
\and
Felix Hormuth Engineering, Goethestr. 17, 69181 Leimen, Germany\label{aff71}
\and
Technical University of Denmark, Elektrovej 327, 2800 Kgs. Lyngby, Denmark\label{aff72}
\and
Cosmic Dawn Center (DAWN), Denmark\label{aff73}
\and
Max-Planck-Institut f\"ur Astronomie, K\"onigstuhl 17, 69117 Heidelberg, Germany\label{aff74}
\and
NASA Goddard Space Flight Center, Greenbelt, MD 20771, USA\label{aff75}
\and
Universit\'e de Gen\`eve, D\'epartement de Physique Th\'eorique and Centre for Astroparticle Physics, 24 quai Ernest-Ansermet, CH-1211 Gen\`eve 4, Switzerland\label{aff76}
\and
Helsinki Institute of Physics, Gustaf H{\"a}llstr{\"o}min katu 2, University of Helsinki, 00014 Helsinki, Finland\label{aff77}
\and
Laboratoire d'etude de l'Univers et des phenomenes eXtremes, Observatoire de Paris, Universit\'e PSL, Sorbonne Universit\'e, CNRS, 92190 Meudon, France\label{aff78}
\and
SKAO, Jodrell Bank, Lower Withington, Macclesfield SK11 9FT, UK\label{aff79}
\and
Centre de Calcul de l'IN2P3/CNRS, 21 avenue Pierre de Coubertin 69627 Villeurbanne Cedex, France\label{aff80}
\and
Universit\"at Bonn, Argelander-Institut f\"ur Astronomie, Auf dem H\"ugel 71, 53121 Bonn, Germany\label{aff81}
\and
INFN-Sezione di Roma, Piazzale Aldo Moro, 2 - c/o Dipartimento di Fisica, Edificio G. Marconi, 00185 Roma, Italy\label{aff82}
\and
Aix-Marseille Universit\'e, CNRS, CNES, LAM, Marseille, France\label{aff83}
\and
Institute of Physics, Laboratory of Astrophysics, Ecole Polytechnique F\'ed\'erale de Lausanne (EPFL), Observatoire de Sauverny, 1290 Versoix, Switzerland\label{aff84}
\and
Dipartimento di Fisica - Sezione di Astronomia, Universit\`a di Trieste, Via Tiepolo 11, 34131 Trieste, Italy\label{aff85}
\and
Institut de F\'{i}sica d'Altes Energies (IFAE), The Barcelona Institute of Science and Technology, Campus UAB, 08193 Bellaterra (Barcelona), Spain\label{aff86}
\and
School of Mathematics and Physics, University of Surrey, Guildford, Surrey, GU2 7XH, UK\label{aff87}
\and
European Space Agency/ESTEC, Keplerlaan 1, 2201 AZ Noordwijk, The Netherlands\label{aff88}
\and
Waterloo Centre for Astrophysics, University of Waterloo, Waterloo, Ontario N2L 3G1, Canada\label{aff89}
\and
Department of Physics and Astronomy, University of Waterloo, Waterloo, Ontario N2L 3G1, Canada\label{aff90}
\and
Perimeter Institute for Theoretical Physics, Waterloo, Ontario N2L 2Y5, Canada\label{aff91}
\and
Universit\'e Paris-Saclay, Universit\'e Paris Cit\'e, CEA, CNRS, AIM, 91191, Gif-sur-Yvette, France\label{aff92}
\and
Space Science Data Center, Italian Space Agency, via del Politecnico snc, 00133 Roma, Italy\label{aff93}
\and
Centre National d'Etudes Spatiales -- Centre spatial de Toulouse, 18 avenue Edouard Belin, 31401 Toulouse Cedex 9, France\label{aff94}
\and
Dipartimento di Fisica e Astronomia "G. Galilei", Universit\`a di Padova, Via Marzolo 8, 35131 Padova, Italy\label{aff95}
\and
INFN-Padova, Via Marzolo 8, 35131 Padova, Italy\label{aff96}
\and
Instituto de F\'isica Te\'orica UAM-CSIC, Campus de Cantoblanco, 28049 Madrid, Spain\label{aff97}
\and
Institut de Recherche en Astrophysique et Plan\'etologie (IRAP), Universit\'e de Toulouse, CNRS, UPS, CNES, 14 Av. Edouard Belin, 31400 Toulouse, France\label{aff98}
\and
Universit\'e St Joseph; Faculty of Sciences, Beirut, Lebanon\label{aff99}
\and
Departamento de F\'isica, FCFM, Universidad de Chile, Blanco Encalada 2008, Santiago, Chile\label{aff100}
\and
Infrared Processing and Analysis Center, California Institute of Technology, Pasadena, CA 91125, USA\label{aff101}
\and
Cosmic Dawn Center (DAWN)\label{aff102}
\and
Niels Bohr Institute, University of Copenhagen, Jagtvej 128, 2200 Copenhagen, Denmark\label{aff103}
\and
Universidad Polit\'ecnica de Cartagena, Departamento de Electr\'onica y Tecnolog\'ia de Computadoras,  Plaza del Hospital 1, 30202 Cartagena, Spain\label{aff104}
\and
European University of Technology EUt+, European Union\label{aff105}
\and
Institute of Space Sciences (ICE, CSIC), Campus UAB, Carrer de Can Magrans, s/n, 08193 Barcelona, Spain\label{aff106}
\and
Institut d'Estudis Espacials de Catalunya (IEEC),  Edifici RDIT, Campus UPC, 08860 Castelldefels, Barcelona, Spain\label{aff107}
\and
Istituto Nazionale di Fisica Nucleare, Sezione di Ferrara, Via Giuseppe Saragat 1, 44122 Ferrara, Italy\label{aff108}
\and
ICSC - Centro Nazionale di Ricerca in High Performance Computing, Big Data e Quantum Computing, Via Magnanelli 2, Bologna, Italy\label{aff109}
\and
INFN-Bologna, Via Irnerio 46, 40126 Bologna, Italy\label{aff110}
\and
Astronomical Observatory of the Autonomous Region of the Aosta Valley (OAVdA), Loc. Lignan 39, I-11020, Nus (Aosta Valley), Italy\label{aff111}
\and
Universidad de La Laguna, Dpto. Astrof\'\i sica, E-38206 La Laguna, Tenerife, Spain\label{aff112}
\and
Universit\"at Innsbruck, Institut f\"ur Astro- und Teilchenphysik, Technikerstr. 25/8, 6020 Innsbruck, Austria\label{aff113}
\and
Oskar Klein Centre for Cosmoparticle Physics, Department of Physics, Stockholm University, Stockholm, SE-106 91, Sweden\label{aff114}
\and
Astrophysics Group, Blackett Laboratory, Imperial College London, London SW7 2AZ, UK\label{aff115}
\and
Dipartimento di Fisica, Sapienza Universit\`a di Roma, Piazzale Aldo Moro 2, 00185 Roma, Italy\label{aff116}
\and
Institut d'Astrophysique de Paris, 98bis Boulevard Arago, 75014, Paris, France\label{aff117}
\and
ICL, Junia, Universit\'e Catholique de Lille, LITL, 59000 Lille, France\label{aff118}
\and
Center for Astrophysics and Cosmology, University of Nova Gorica, Nova Gorica, Slovenia\label{aff119}
\and
Department of Astronomy, Indiana University, 727 East Third Street, Bloomington, IN 47405, USA\label{aff120}
\and
CERCA/ISO, Department of Physics, Case Western Reserve University, 10900 Euclid Avenue, Cleveland, OH 44106, USA\label{aff121}
\and
Departamento de F{\'\i}sica Fundamental. Universidad de Salamanca. Plaza de la Merced s/n. 37008 Salamanca, Spain\label{aff122}
\and
Universiteit Gent, Department of Physics and Astronomy, Proeftuinstraat 86 N3, 9000 Ghent, Belgium
\label{aff123}
\and
Konkoly Observatory, HUN-REN CSFK, MTA Centre of Excellence, Budapest, Konkoly Thege Mikl\'os {\'u}t 15-17. H-1121, Hungary\label{aff124}
\and
MTA-CSFK Lend\"ulet Large-Scale Structure Research Group, Konkoly-Thege Mikl\'os \'ut 15-17, H-1121 Budapest, Hungary\label{aff125}
\and
Universit\'e de Strasbourg, CNRS, Observatoire astronomique de Strasbourg, UMR 7550, 67000 Strasbourg, France\label{aff126}
\and
Scuola Superiore Meridionale, Via Mezzocannone 4, 80138, Napoli, Italy\label{aff127}
\and
California Institute of Technology, 1200 E California Blvd, Pasadena, CA 91125, USA\label{aff128}
\and
Instituto de Astrof\'isica e Ci\^encias do Espa\c{c}o, Faculdade de Ci\^encias, Universidade de Lisboa, Tapada da Ajuda, 1349-018 Lisboa, Portugal\label{aff129}
\and
Kapteyn Astronomical Institute, University of Groningen, PO Box 800, 9700 AV Groningen, The Netherlands\label{aff130}
\and
Institute for Astronomy, University of Hawaii, 2680 Woodlawn Drive, Honolulu, HI 96822, USA\label{aff131}
\and
Department of Mathematics and Physics E. De Giorgi, University of Salento, Via per Arnesano, CP-I93, 73100, Lecce, Italy\label{aff132}
\and
INFN, Sezione di Lecce, Via per Arnesano, CP-193, 73100, Lecce, Italy\label{aff133}
\and
INAF-Sezione di Lecce, c/o Dipartimento Matematica e Fisica, Via per Arnesano, 73100, Lecce, Italy\label{aff134}
\and
Departamento F\'isica Aplicada, Universidad Polit\'ecnica de Cartagena, Campus Muralla del Mar, 30202 Cartagena, Murcia, Spain\label{aff135}
\and
Instituto de F\'isica de Cantabria, Edificio Juan Jord\'a, Avenida de los Castros, 39005 Santander, Spain\label{aff136}
\and
Univ. Grenoble Alpes, CNRS, IPAG, 38000 Grenoble, France\label{aff137}
\and
Dipartimento di Fisica "E. Pancini", Universita degli Studi di Napoli Federico II, Via Cinthia 6, 80126, Napoli, Italy\label{aff138}
\and
INFN Sezione di Napoli, Universita degli Studi di Napoli "Federico II", Compl. Univ. di Monte S. Angelo, Edificio G, Via Cinthia, I-80126, Napoli, Italy\label{aff139}
\and
Universit\'e Paris Cit\'e, CNRS, Astroparticule et Cosmologie, 75013 Paris, France\label{aff140}
\and
Institut d'Astrophysique de Paris, UMR 7095, CNRS, and Sorbonne Universit\'e, 98 bis boulevard Arago, 75014 Paris, France\label{aff141}
\and
Department of Computer Science, Aalto University, PO Box 15400, Espoo, FI-00 076, Finland\label{aff142}
\and
University of Applied Sciences and Arts of Northwestern Switzerland, School of Computer Science, 5210 Windisch, Switzerland\label{aff143}
\and
DARK, Niels Bohr Institute, University of Copenhagen, Jagtvej 155, 2200 Copenhagen, Denmark\label{aff144}
\and
Department of Physics, Centre for Extragalactic Astronomy, Durham University, South Road, Durham, DH1 3LE, UK\label{aff145}
\and
Astrophysics Research Centre, University of KwaZulu-Natal, Westville Campus, Durban 4041, South Africa\label{aff146}
\and
School of Mathematics, Statistics \& Computer Science, University of KwaZulu-Natal, Westville Campus, Durban 4041, South Africa\label{aff147}
\and
Department of Physics and Astronomy, Vesilinnantie 5, University of Turku, 20014 Turku, Finland\label{aff148}
\and
Serco for European Space Agency (ESA), Camino bajo del Castillo, s/n, Urbanizacion Villafranca del Castillo, Villanueva de la Ca\~nada, 28692 Madrid, Spain\label{aff149}
\and
Van Swinderen Institute, University of Groningen, Nijenborgh 3, 9747 AG Groningen, The Netherlands\label{aff150}
\and
Department of Physics and Astronomy, University of the Western Cape, Bellville, Cape Town, 7535, South Africa\label{aff151}
\and
Department of Physics and Helsinki Institute of Physics, Gustaf H\"allstr\"omin katu 2, University of Helsinki, 00014 Helsinki, Finland\label{aff152}
\and
Ohio University, Physics \& Astronomy Department,1 Ohio University, Athens, OH 45701, USA\label{aff153}
\and
Institute for Theoretical Particle Physics and Cosmology (TTK), RWTH Aachen University, 52056 Aachen, Germany\label{aff154}
\and
IRFU, CEA, Universit\'e Paris-Saclay 91191 Gif-sur-Yvette Cedex, France\label{aff155}
\and
INAF-Osservatorio Astrofisico di Arcetri, Largo E. Fermi 5, 50125, Firenze, Italy\label{aff156}
\and
Centro de Astrof\'{\i}sica da Universidade do Porto, Rua das Estrelas, 4150-762 Porto, Portugal\label{aff157}
\and
HE Space for European Space Agency (ESA), Camino bajo del Castillo, s/n, Urbanizacion Villafranca del Castillo, Villanueva de la Ca\~nada, 28692 Madrid, Spain\label{aff158}
\and
Center for Frontier Science, Chiba University, 1-33 Yayoi-cho, Inage-ku, Chiba 263-8522, Japan\label{aff159}
\and
Department of Physics, Graduate School of Science, Chiba University, 1-33 Yayoi-Cho, Inage-Ku, Chiba 263-8522, Japan\label{aff160}
\and
Dipartimento di Fisica ``G. Occhialini", Universit\`a degli Studi di Milano Bicocca, Piazza della Scienza 3, 20126 Milano, Italy\label{aff161}
\and
Department of Astrophysics, University of Zurich, Winterthurerstrasse 190, 8057 Zurich, Switzerland\label{aff162}
\and
Department of Mathematics and Physics, Roma Tre University, Via della Vasca Navale 84, 00146 Rome, Italy\label{aff163}
\and
Theoretical astrophysics, Department of Physics and Astronomy, Uppsala University, Box 524, 751 20 Uppsala, Sweden\label{aff164}
\and
Mathematical Institute, University of Leiden, Einsteinweg 55, 2333 CA Leiden, The Netherlands\label{aff165}
\and
School of Physical Sciences, The Open University, Milton Keynes, MK7 6AA, UK\label{aff166}
\and
School of Physics \& Astronomy, University of Southampton, Highfield Campus, Southampton SO17 1BJ, UK\label{aff167}
\and
Institute of Astronomy, University of Cambridge, Madingley Road, Cambridge CB3 0HA, UK\label{aff168}
\and
Department of Astrophysical Sciences, Peyton Hall, Princeton University, Princeton, NJ 08544, USA\label{aff169}
\and
Fakult\"at f\"ur Physik, Universit\"at Bielefeld, Postfach 100131, 33501 Bielefeld, Germany\label{aff170}
\and
Space physics and astronomy research unit, University of Oulu, Pentti Kaiteran katu 1, FI-90014 Oulu, Finland\label{aff171}
\and
Department of Physics and Astronomy, Lehman College of the CUNY, Bronx, NY 10468, USA\label{aff172}
\and
American Museum of Natural History, Department of Astrophysics, New York, NY 10024, USA\label{aff173}
\and
Ruhr University Bochum, Faculty of Physics and Astronomy, Astronomical Institute (AIRUB), German Centre for Cosmological Lensing (GCCL), 44780 Bochum, Germany\label{aff174}}    
\clearpage

%
%
%
%


%
%
\abstract{Strong gravitational-lensing systems with multiple source planes are powerful tools for probing the stellar and dark matter content of galaxies. Lens modelling can measure the relative Einstein radii of the two source planes, which encode a ratio of angular-diameter distances determined by the lens and source redshifts and the dark energy equation of state, independently of the Hubble constant. Galaxy-scale double-source-plane lenses (DSPLs) are extremely rare. We present a sample of \sampleprovisional{84} galaxy-scale DSPL candidates in Euclid Data Release~1 (DR1), of which \sampleprovisional{80} are newly identified. These systems were selected using machine-learning lens-finding models followed by visual inspection by citizen scientists and experts. We then used lens modelling to test the DSPL interpretation. Among our lens candidates, \sampleprovisional{11} form the `Gold' sample: each shows two sets of multiple images at different radii, and the observed configuration is reproduced by a DSPL lens model. The remaining \sampleprovisional{73} `Silver' candidates have at least two lensed arcs from visual inspection, but either lack multiple images for at least one arc, have a best-fitting DSPL distance ratio close to the single-source-plane limit, remain compatible with a single-plane model with multiple sources, or have not yet been modelled. Even if all \sampleprovisional{84} candidates are DSPLs, the sample falls below the approximately 90 systems expected from extrapolating the Euclid Quick Data Release 1 (Q1) yield and the approximately 200 obtained by scaling the LensPop forecast to the DR1 area, indicating that many DSPLs remain undetected in DR1. Spectroscopy or higher-resolution imaging is required before we can be confident that the Silver candidates are DSPLs.}
%
%
%
%
   \titlerunning{Double-source-plane lens candidates}
   \authorrunning{Euclid Collaboration: Tian Li et al.}
   
   \maketitle
\fi
%
%
%
%
   
\section{\label{sc:Intro}Introduction}

Strong gravitational lensing occurs when the path of light from a distant background source is apparently bent by the gravity of a foreground massive object that lies nearly along the same line of sight. In this situation, the source can appear as multiple images, extended arcs, or, when the alignment is almost perfect, an Einstein ring \citep{Einstein1936, Zwicky1937a, Zwicky1937b}. In some lensing systems, more than one background source is located behind the same foreground lens, and these sources are at different redshifts. Such multiple source planes can produce separate sets of lensed images, often seen as Einstein rings at different angular radii. These systems are usually called compound lenses or DSPLs.

\begin{figure*}
    \centering
    \makebox[\textwidth][c]{%
        \includegraphics[height=0.282\textwidth]{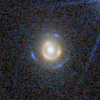}%
        \includegraphics[height=0.282\textwidth]{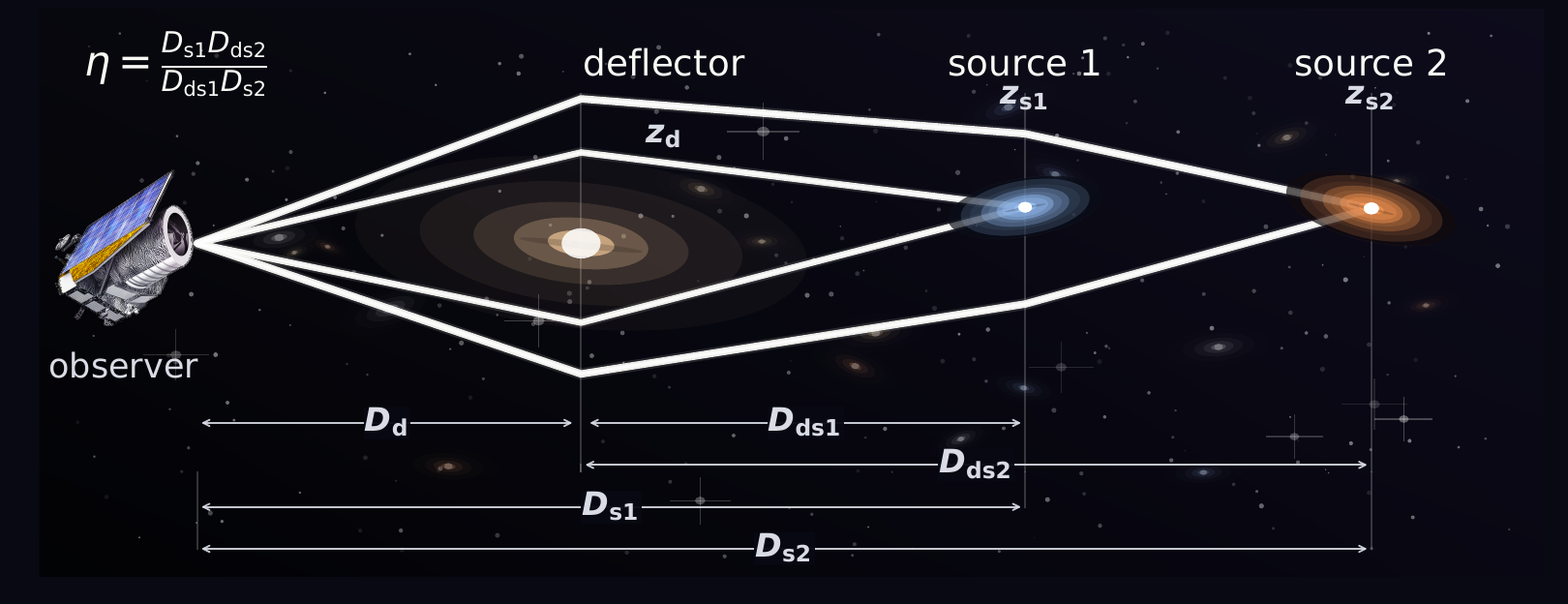}%
    }
    \caption{Example and geometry of a DSPL. \emph{Left}: a DSPL discovered in DR1, with the \JE, \YE, and \IE bands mapped to the red, green, and blue channels, respectively. \emph{Right}: schematic lensing geometry showing the observer, foreground deflector at redshift \(z_{\mathrm{d}}\), and two background sources at redshifts \(z_{\mathrm{s1}}\) and \(z_{\mathrm{s2}}\). Euclid spacecraft image: ESA/C.~Carreau.}
    \label{fig:DSPL_example}
\end{figure*}

In a strong-lensing configuration, the angular scale of the Einstein ring, usually written as $\theta_{\sfont{E}}$, is called the Einstein radius. This quantity depends on both the projected mass distribution of the lens and the relevant angular-diameter distances. For lenses with a single source plane, detailed lens modelling can constrain the local logarithmic density slope by fitting the deflection field in the region where the multiple images or arcs form \citep[e.g.,][]{Suyu2010, Birrer2018, Nightingale2021, Galan2022}. However, such inferences are affected by systematic errors in lens modelling. One important example is the mass-sheet degeneracy \citep[]{Falco1985, Schneider:2013sxa}. In this degeneracy, adding a uniform mass sheet and rescaling the convergence can change the inferred mass-profile shape and the intrinsic size of the source, while the lensing observables remain the same. In DSPL systems, the presence of a second source plane can help break this degeneracy, provided that the mass of the first source is negligible and that the background cosmology is fixed \citep{Brada2004}. An example image and the lensing geometry of a DSPL system are shown in Fig.~\ref{fig:DSPL_example}.

The presence of two Einstein rings at different radii makes DSPLs powerful probes of lens-galaxy mass distributions. The second ring provides an additional enclosed-mass measurement, helping to disentangle the distributions of dark and luminous matter and improving sensitivity to dark matter substructure \ifdefined\PaperSixThesisMode\citep{Tian2026, Enzi2024}\else\citep{Li2026jackpot, Enzi2024}\fi; studies of SDSS J0946+1006 illustrate this potential, although the claimed over-concentrated subhalo remains contested \ifdefined\PaperSixThesisMode\citep{vegetti2010detection, Minor:2020hic, Ballard:2023fgi, Despali:2024ihn, Enzi2024, Qiuhan2025}\else\citep{vegetti2010detection, Minor:2020hic, Ballard:2023fgi, Despali:2024ihn, Enzi2024, He2025}\fi. The relative Einstein radii of the two sources also constrain the DSPL distance ratio $\eta$ (Eq.~\ref{eq:eta}), whose lensing geometry is illustrated in Fig.~\ref{fig:DSPL_example}, providing a probe of the dark energy equation of state that is independent of the Hubble constant \ifdefined\PaperSixThesisMode\citep[e.g.,][]{gavazzi2008, collett2012, Collett2014, Johnson2025, Bowden2025}\else\citep[e.g.,][]{Gavazzi2008, collett2012, Collett2014, Johnson2025, Bowden2025}\fi. However, a generalised mass-sheet transformation can weaken these cosmographic constraints \citep{Schneider2014}, and breaking this degeneracy with stellar kinematics requires unbiased measurements and accurate assumptions about the lens mass profile, both of which are very challenging.

Galaxy-scale strong lenses with multiple background sources remain rare, because the probability of lensing two independent source planes scales approximately as $\theta_{\sfont{E}}^4$, compared to $\theta_{\sfont{E}}^2$ for a single source plane. Among the fewer than 1000 strong lenses discovered before \Euclid began its survey, fewer than ten were confirmed DSPLs, and only four of these were suitable for cosmographic analyses \ifdefined\PaperSixThesisMode\citep{Li2025b}\else\citep{Q1-SP054}\fi. To our knowledge, the following studies account for the full published sample of galaxy-scale DSPLs and DSPL candidates available before this work. Early examples include the Jackpot lens \ifdefined\PaperSixThesisMode\citep{gavazzi2008}\else\citep{Gavazzi2008}\fi~and the Eye of Horus \citep{Tanaka2016}, while systems with sources at similar Einstein radii include J1721+8842 \citep{Dux2024}, the Dark Energy Survey (DES) lens DES~J0408$-$5354 \citep{2017ApJ...838L..15L}, and the Cosmic Horseshoe \citep{2007ApJ...671L...9B}. Recent wide-field lens searches are now beginning to enlarge this sample. The ASTRO 3D Galaxy Evolution with Lenses (AGEL) survey reports six galaxy-scale DSPL candidates, including AGEL150745+052256 and AGEL035346$-$170639, which are used for DSPL cosmography \citep{Barone2026, Bowden2025}.

Most pre-\Euclid DSPLs were initially selected by galaxy--galaxy lens-finding programmes using ground-based spectroscopy or imaging and were subsequently observed with the \textit{Hubble} Space Telescope (HST), whose higher spatial resolution revealed DSPL features that were not discernible in the ground-based data. The search is more direct with \Euclid: its VIS imaging has a point-spread function (PSF) full width at half maximum (FWHM) of approximately \ang{;;0.16} \citep{EuclidSkyVIS}, allowing multiple-source-plane features to be recognised in the discovery data themselves, while its three near-infrared bands (\YE, \JE, and \HE) help identify additional arcs from high-redshift sources \citep{EuclidSkyNISP}. Lens modelling can then test and support the DSPL interpretation. The Euclid Quick Data Release 1 (Q1) reports three robust galaxy-scale DSPL candidates over 63\,deg$^2$, with one additional object whose DSPL nature remains uncertain.

In this work, we searched for DSPLs in the 1900\,deg$^2$ covered by Euclid Data Release 1 (DR1; \citealt{DR1cite}; Euclid Collaboration: Aussel et al., in prep.; Euclid Collaboration: McCracken et al., in prep.). We used the strong-lensing discovery engine, a structured workflow that integrates machine learning with citizen-science classifications followed by expert grading, as described in the ECLIPSE catalogue and discovery-pipeline papers (Euclid Collaboration: Lines et al., in prep.; Euclid Collaboration: Holloway et al., in prep.). Candidate systems were selected during visual inspection of the highest-grade lens candidates. The single-plane lens-modelling results presented by Euclid Collaboration: Nightingale et al. (in prep.) further reveal several systems with additional arcs, which we included in our candidate list. Finally, we used the DSPL modelling described in Sect.~\ref{sec:modelling} to measure $\eta$ and assess the DSPL nature of the modelled candidates. To our knowledge, this is the largest search for DSPLs conducted to date. In total, we identified \sampleprovisional{84} galaxy-scale DSPL candidates, of which \sampleprovisional{80} are newly identified; three are also reported in Q1 \ifdefined\PaperSixThesisMode\citep{Li2025b}\else\citep{Q1-SP054}\fi, and one, DES~J0408$-$5354, was already known from DES \citep{2017ApJ...838L..15L}. Of these, \sampleprovisional{11} form our Gold sample, for which the morphology and lens models strongly support a DSPL interpretation, although spectroscopic redshifts are still required to confirm it. The remaining \sampleprovisional{73} form our Silver sample, which additionally requires higher-resolution imaging and more detailed lens modelling to establish its DSPL nature.

The paper is organised as follows. Section~\ref{sec:modelling} introduces the lensing formalism and the modelling strategy used to interpret the candidates. Section~\ref{sec:result} presents the Gold and Silver DSPL samples and the lens-model results. Section~\ref{sec:discussion} discusses several ambiguous objects and the limitations of the current classification. Section~\ref{sec:forecast} compares the number of DSPLs found in DR1 with expectations from previous forecasts. Finally, Section~\ref{sec:conclusion} summarises our conclusions. The appendices present the Gold lens-model reconstructions and list the official designations and sky coordinates of the Silver candidates.

\section{Modelling double-source-plane lenses}
\label{sec:modelling}

\subsection{Background theory}

\begin{figure*}
    \centering
    \begin{tabular}{cccc}
        \numberedscalebarimage{1}{0.22}{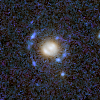}{20}{2}
        \numberedimage{2}{0.22}{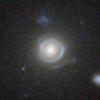}
        \numberedimage{3}{0.22}{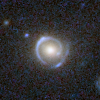}
        \numberedimage{4}{0.22}{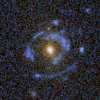}\\
        \numberedimage{5}{0.22}{figures/021_DSPL_RA73_572905DECNEG34_164155_VISYJ_Tian_color_mtf.jpg}
        \numberedimage{6}{0.22}{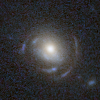}
        \numberedimage{7}{0.22}{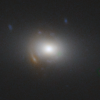}
        \numberedimage{8}{0.22}{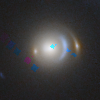}\\
        \numberedimage{9}{0.22}{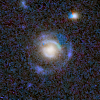}
        \numberedimage{10}{0.22}{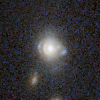}
        \numberedimage{11}{0.22}{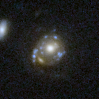}
    \end{tabular}
    \caption{Colour images of all \sampleprovisional{11} Gold DSPL candidates (see Sect.~\ref{sec:result} for the Gold and Silver classification). The colour composites map the \JE, \YE, and \IE bands to the red, green, and blue channels, respectively. The numbers in the upper-left corners correspond to the entries in Table~\ref{tab:tier1_result}, which are ordered by increasing right ascension. Each cutout is \ang{;;10} on a side; the scale bar and compass in the first panel indicate \ang{;;2} and the orientation (north up, east to the left), which is the same for all panels.}
    \label{fig:Tier1image}
\end{figure*}

In this section, we present the theory, conventions, and notation used across the paper to describe the adopted lens-modelling techniques.
For a lensed image at position $\boldsymbol{\theta}$, the scaled deflection angle of a lens galaxy $\boldsymbol{\alpha}(\boldsymbol{\theta})$ is related to its deflection potential $\psidp$ via
\begin{equation}
\boldsymbol{\alpha}(\boldsymbol{\theta})=\Nabla \psidp(\boldsymbol{\theta})\;,
\end{equation}
and the relation between the deflection potential and the lensing convergence is
\begin{equation}
\kappa(\boldsymbol{\theta})=\frac{1}{2}\,\nabla^2 \psidp(\boldsymbol{\theta})\;,
\end{equation}
where convergence is defined as
\begin{equation}
\kappa(\boldsymbol{\theta}) \equiv \frac{\Sigma(\Dd\boldsymbol{\theta})}{\Sigcr}\;.
\end{equation}
The convergence is the lens surface mass density $\Sigma$ normalised by the critical lensing surface density $\Sigcr$, which is
\begin{equation}
\Sigcr \equiv \frac{c^2}{4\pi\,\GN}\,\frac{\Ds}{\Dd\,\Dds}\;,
\end{equation}
where \Dd{} is the angular-diameter distance to the deflector, \Ds{} is the angular-diameter distance to the source, \Dds{} is the angular-diameter distance from the deflector to the source, $c$ is the speed of light, and \GN{} is Newton's gravitational constant.

In a DSPL system, the lens equation of the first source plane can be written as
\begin{equation}
\boldsymbol{\beta}_1 = \boldsymbol{\theta} - \boldsymbol{\alpha}_1(\boldsymbol{\theta})\;,
\end{equation}
where \({\boldsymbol{\alpha}_1}\) is the scaled deflection angle of the primary deflector for source 1 at the image position $\boldsymbol{\theta}$, and \(\boldsymbol{\beta}_1\) is the unlensed angular position of source 1. The lens equation of the second source plane is then
\begin{equation}
\boldsymbol{\beta}_2 = \boldsymbol{\theta} - \eta\,\boldsymbol{\alpha}_1(\boldsymbol{\theta}) - \boldsymbol{\alpha}_{\mathrm{s1}}[\boldsymbol{\theta} - \boldsymbol{\alpha}_1(\boldsymbol{\theta})]\;,
\end{equation}
where $\boldsymbol{\beta}_2$ is the unlensed angular position of source 2, and $\boldsymbol{\alpha}_{\mathrm{s1}}$ is the deflection angle of the mass on the first source plane. Here, \(\eta\) is the ratio of the angular-diameter distances between the different redshift planes,
\begin{equation}
\label{eq:eta}
\eta = \frac{D_{\mathrm{s1}}\,D_{\mathrm{ds2}}}{D_{\mathrm{ds1}}\,D_{\mathrm{s2}}}\;,
\end{equation}
where the subscripts \(\mathrm{s1}\) and \(\mathrm{s2}\) denote source 1 and source 2, respectively, and \(\mathrm{d}\) denotes the deflector. In a simple case for which we assume that the lens galaxy has a singular isothermal sphere (SIS) profile, the parameter \(\eta\) is just the ratio of the two Einstein radii.

We modelled the lenses using an elliptical power-law (EPL) mass model \citep{2015A&A...580A..79T} with an external shear component. The convergence of the EPL model can be parametrised as
\begin{equation}
\kappa_{\sfont{EPL}}(\rho, \gamma_{\sfont{EPL}}, q)=\frac{3-\gamma_{\sfont{EPL}}}{2}\,\left(\frac{\theta_{\sfont{E}}}{\rho}\right)^{\gamma_{\sfont{EPL}}-1}\;.
\end{equation}
Here we choose a Cartesian coordinate system in which the $x$- and $y$-axes are aligned with the principal axes of the EPL mass ellipse. In this coordinate system, $\rho^2=x^2 q+y^2/q$ is the squared elliptical radius, $q$ is the axis ratio, $\gamma_{\sfont{EPL}}$ is the power-law index of the density slope, and $\theta_{\sfont{E}}$ is the Einstein radius. We use $\theta_{\sfont{E},1}$ to represent the Einstein radius of the primary lens with respect to source 1, which is the free Einstein-radius parameter of the EPL model. The Einstein radius of the same lens with respect to source 2 (with source 1 absent), $\theta_{\sfont{E},2}$, is not a free parameter: it is fixed by $\theta_{\sfont{E},1}$, $\eta$, and $\gamma_{\sfont{EPL}}$, and serves only as a derived reference quantity. The mass of source 1 is not neglected: we model it as an SIS, whose deflection is $\boldsymbol{\alpha}_{\mathrm{s1}}$ in the lens equation of the second source plane, and we parametrise it by $\theta_{\sfont{E},\sfont{SIS}}$, the Einstein radius of source 1 for light from source 2. Both $\theta_{\sfont{E},1}$ and $\theta_{\sfont{E},\sfont{SIS}}$ are free parameters, with the priors listed in Table~\ref{tab:default_priors}. The external shear approximates the tidal perturbation from mass outside the explicitly modelled deflector, such as nearby galaxies or line-of-sight structure. For the stochastic variational inference (SVI; \citealt{WingateWeber2013}) models used in this work, we parametrised the external shear using the Cartesian components $\gamma_{\sfont{ext},1}$ and $\gamma_{\sfont{ext},2}$ and assigned each component an independent uniform prior. This choice is not rotationally invariant and therefore introduces a preference in the shear direction. We use it at the SVI stage because the Cartesian components vary continuously and allow us to find a good approximate lens model quickly. When the shear is sampled using its strength and position angle, the angle is periodic but the parameter interval has two boundaries. Two physically equivalent directions can then lie on opposite sides of the interval, so a gradient-based optimiser cannot move smoothly between them and may become stuck at a boundary. For rigorous posterior sampling with Markov chain Monte Carlo methods such as Hamiltonian Monte Carlo, we will instead sample the shear strength and position angle with an isotropic prior and treat the angular periodicity explicitly.

\begin{figure*}[!t]
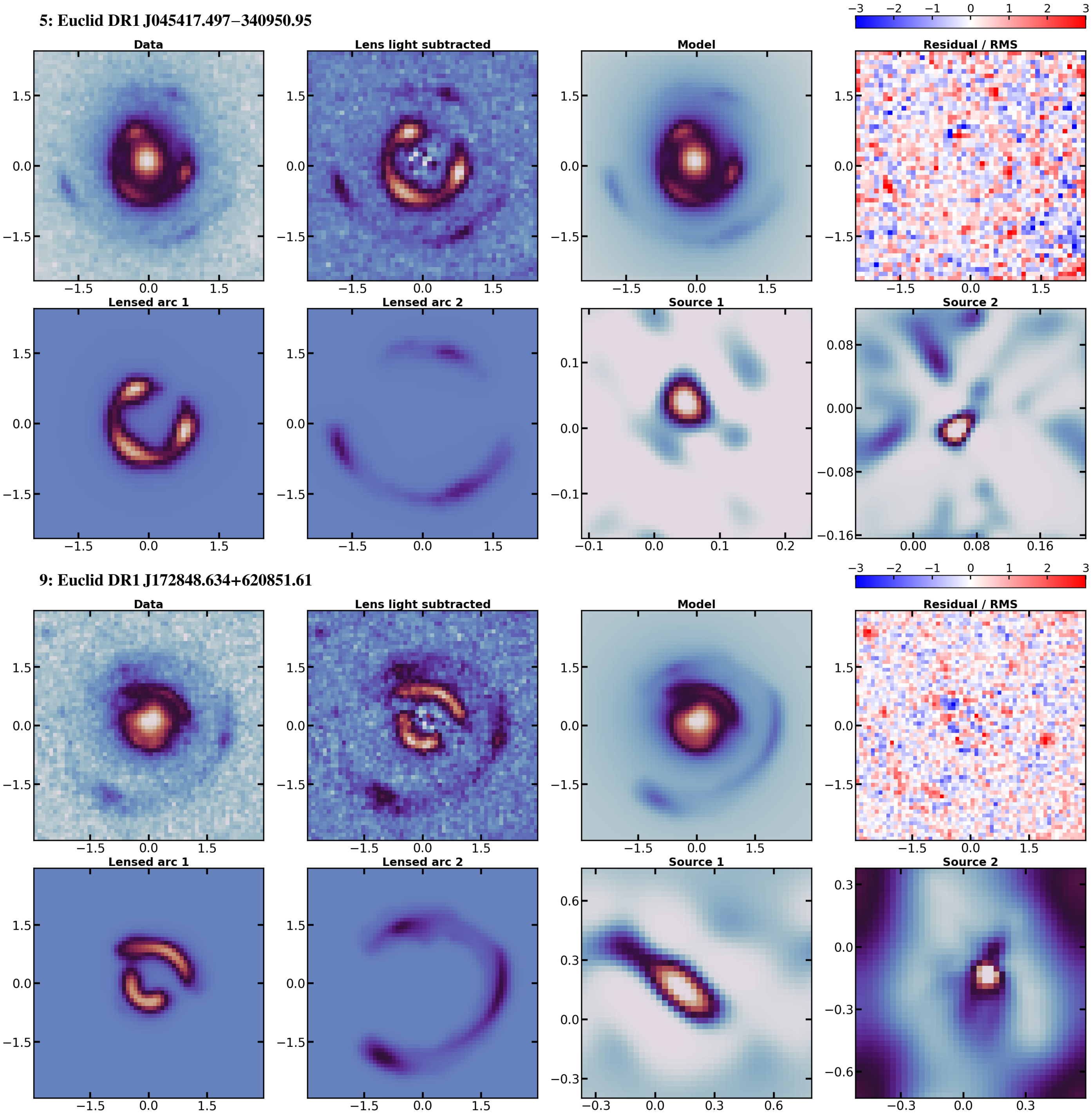

    \centering
    \labeledmodelimage{5: Euclid DR1\,J045417.497$-$340950.95}{1}{figures/DSPL_RA73_572905DECNEG34_164155_chain01_pixelated_min_loss_style4}
    \par\vspace{0.6em}
    \labeledmodelimage{9: Euclid DR1\,J172848.634+620851.61}{1}{figures/DSPL_RA262_202643DEC62_147672_chain04_pixelated_min_loss_style4}
    \caption{Example lens models of two Gold DSPL candidates. Each row shows a different lens system; from left to right, the panels show the observed cutout, lens-light-subtracted data, reconstructed model, normalised residuals, the separated lensed arcs from the two source planes, and the corresponding pixelated source reconstructions. The scale of all coordinate axes is arcseconds.}
    \label{fig:tier1example}
\end{figure*}

\begin{figure*}[p]
    \centering
    \begingroup
    \setlength{\tabcolsep}{1pt}
    \renewcommand{\arraystretch}{0.55}
    \begin{tabular}{cccccccc}
        \numberedscalebarimage[trim=50.5bp 51bp 50.5bp 51bp,clip]{1}{0.121}{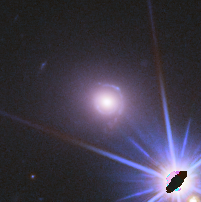}{20}{2} &
        \numberedimage[trim=51bp 50.5bp 51bp 50.5bp,clip]{2}{0.121}{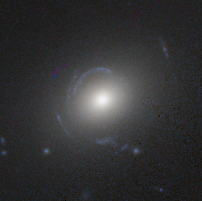} &
        \numberedimage[trim=51bp 50.5bp 51bp 50.5bp,clip]{3}{0.121}{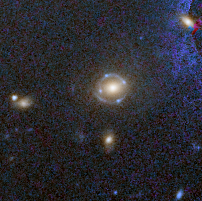} &
        \numberedimage{4}{0.121}{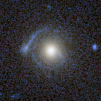} &
        \numberedimage{5}{0.121}{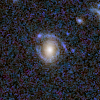} &
        \numberedimage[trim=50.5bp 51bp 50.5bp 51bp,clip]{6}{0.121}{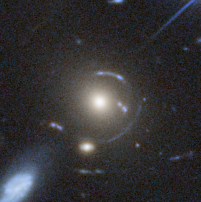} &
        \numberedimage{7}{0.121}{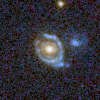} &
        \numberedimage{8}{0.121}{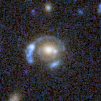}\\[-0.4pt]
        \numberedimage{9}{0.121}{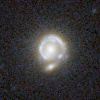} &
        \numberedimage{10}{0.121}{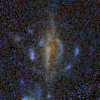} &
        \numberedimage{11}{0.121}{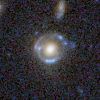} &
        \numberedimage{12}{0.121}{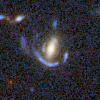} &
        \numberedimage{13}{0.121}{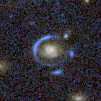} &
        \numberedimage{14}{0.121}{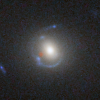} &
        \numberedimage{15}{0.121}{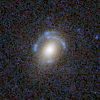} &
        \numberedimage{16}{0.121}{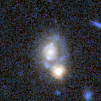}\\[-0.4pt]
        \numberedimage{17}{0.121}{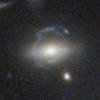} &
        \numberedimage[trim=50.5bp 50.5bp 50.5bp 50.5bp,clip]{18}{0.121}{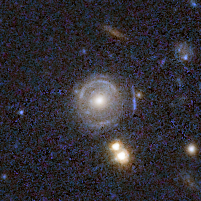} &
        \numberedimage{19}{0.121}{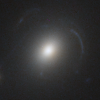} &
        \numberedimage{20}{0.121}{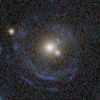} &
        \numberedimage[trim=50.5bp 50.5bp 50.5bp 50.5bp,clip]{21}{0.121}{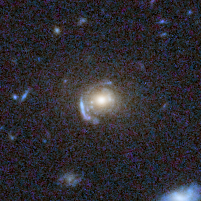} &
        \numberedimage{22}{0.121}{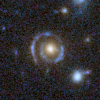} &
        \numberedimage{23}{0.121}{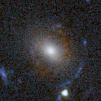} &
        \numberedimage{24}{0.121}{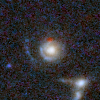}\\[-0.4pt]
        \numberedimage{25}{0.121}{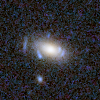} &
        \numberedimage{26}{0.121}{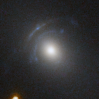} &
        \numberedimage{27}{0.121}{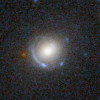} &
        \numberedimage{28}{0.121}{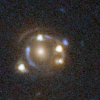} &
        \numberedimage{29}{0.121}{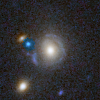} &
        \numberedimage{30}{0.121}{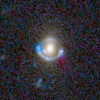} &
        \numberedimage{31}{0.121}{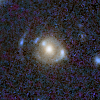} &
        \numberedimage{32}{0.121}{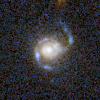}\\[-0.4pt]
        \numberedimage{33}{0.121}{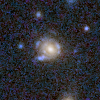} &
        \numberedimage{34}{0.121}{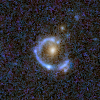} &
        \numberedimage{35}{0.121}{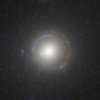} &
        \numberedimage{36}{0.121}{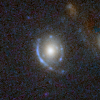} &
        \numberedimage{37}{0.121}{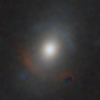} &
        \numberedimage{38}{0.121}{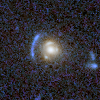} &
        \numberedimage{39}{0.121}{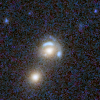} &
        \numberedimage{40}{0.121}{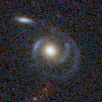}\\[-0.4pt]
        \numberedimage{41}{0.121}{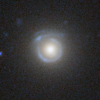} &
        \numberedimage{42}{0.121}{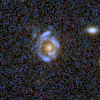} &
        \numberedimage{43}{0.121}{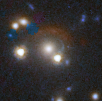} &
        \numberedimage{44}{0.121}{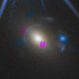} &
        \numberedimage[trim=50.5bp 50.5bp 50.5bp 50.5bp,clip]{45}{0.121}{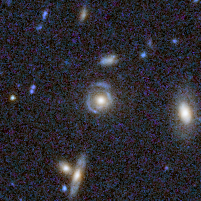} &
        \numberedimage{46}{0.121}{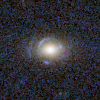} &
        \numberedimage{47}{0.121}{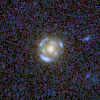} &
        \numberedimage[trim=50.5bp 50.5bp 50.5bp 50.5bp,clip]{48}{0.121}{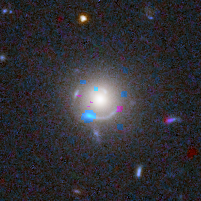}\\[-0.4pt]
        \numberedimage{49}{0.121}{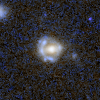} &
        \numberedimage{50}{0.121}{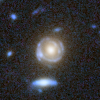} &
        \numberedimage{51}{0.121}{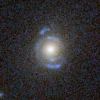} &
        \numberedimage{52}{0.121}{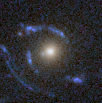} &
        \numberedimage{53}{0.121}{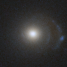} &
        \numberedimage{54}{0.121}{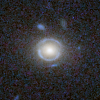} &
        \numberedimage{55}{0.121}{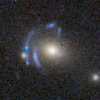} &
        \numberedimage{56}{0.121}{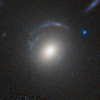}\\[-0.4pt]
        \numberedimage{57}{0.121}{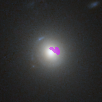} &
        \numberedimage{58}{0.121}{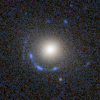} &
        \numberedimage{59}{0.121}{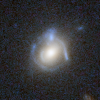} &
        \numberedimage{60}{0.121}{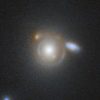} &
        \numberedimage{61}{0.121}{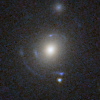} &
        \numberedimage[trim=51bp 51bp 51bp 51bp,clip]{62}{0.121}{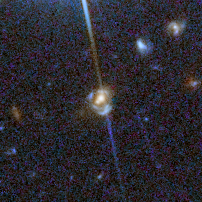} &
        \numberedimage{63}{0.121}{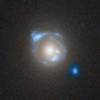} &
        \numberedimage{64}{0.121}{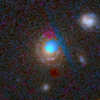}\\[-0.4pt]
        \numberedscalebarimage{65}{0.121}{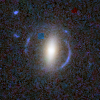}{20}{2} &
        \numberedimage{66}{0.121}{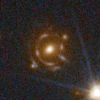} &
        \numberedimage{67}{0.121}{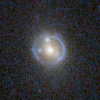} &
        \numberedimage{68}{0.121}{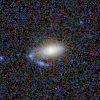} &
        \numberedimage{69}{0.121}{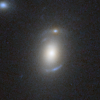} &
        \numberedimage{70}{0.121}{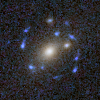} &
        \numberedimage{71}{0.121}{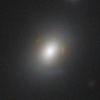} &
        \numberedimage{72}{0.121}{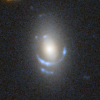}\\[-0.4pt]
        \multicolumn{8}{c}{\numberedimage{73}{0.121}{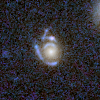}}
    \end{tabular}
    \endgroup
    \caption{Colour images of all \sampleprovisional{73} Silver DSPL candidates, shown with the same setup as Fig.~\ref{fig:Tier1image}. The colour composites map the \JE, \YE, and \IE bands to the red, green, and blue channels, respectively. For panel 37, the \IE, \YE, and \HE bands are instead mapped to the red, green, and blue channels, respectively. The numbers in the upper-left corners correspond to the Silver entries in Table~\ref{tab:tier2_ra_dec}, which are ordered by increasing right ascension. All panels are approximately \ang{;;10} on a side, except panels 44 and 53, which use the native \ang{;;7.9} and \ang{;;6.7} modelling cutouts, respectively; the scale bar and compass in the first panel indicate \ang{;;2} and the orientation (north up, east to the left), which is the same for all panels.}
    \label{fig:Tier2image}
\end{figure*}

\begin{figure*}
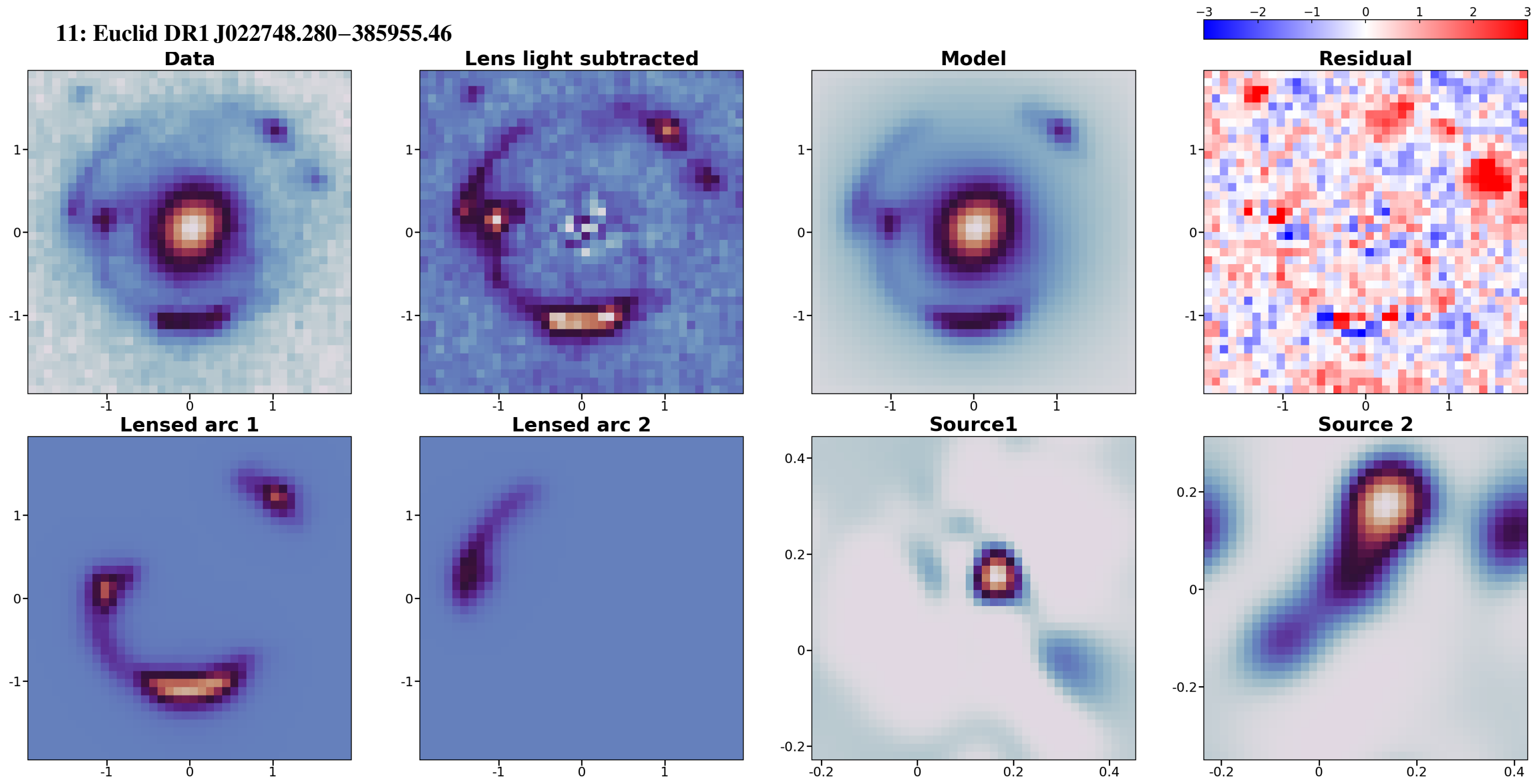

    \centering
    \labeledmodelimage{11: Euclid DR1\,J022748.280$-$385955.46}{1}{figures/DSPL_RA36_951169DECNEG38_998741_chain02_pixelated_min_loss_style4}
    \caption{Same panel layout as Fig.~\ref{fig:tier1example}, but for Silver DSPL candidate 11 (Euclid DR1\,J022748.280$-$385955.46). At least one source does not produce a clear multiple-image configuration, so the object is kept in the Silver sample.}
    \label{fig:Tier2example}
\end{figure*}

\begin{figure*}
    \centering
    \labeledmodelimagesingleplane{47: Euclid DR1\,J053233.630$-$371545.77 }{1}{figures/RA83p140125_DECm37p262715_chain01_pixelated_min_loss_style4}
    \includegraphics[width = \linewidth]{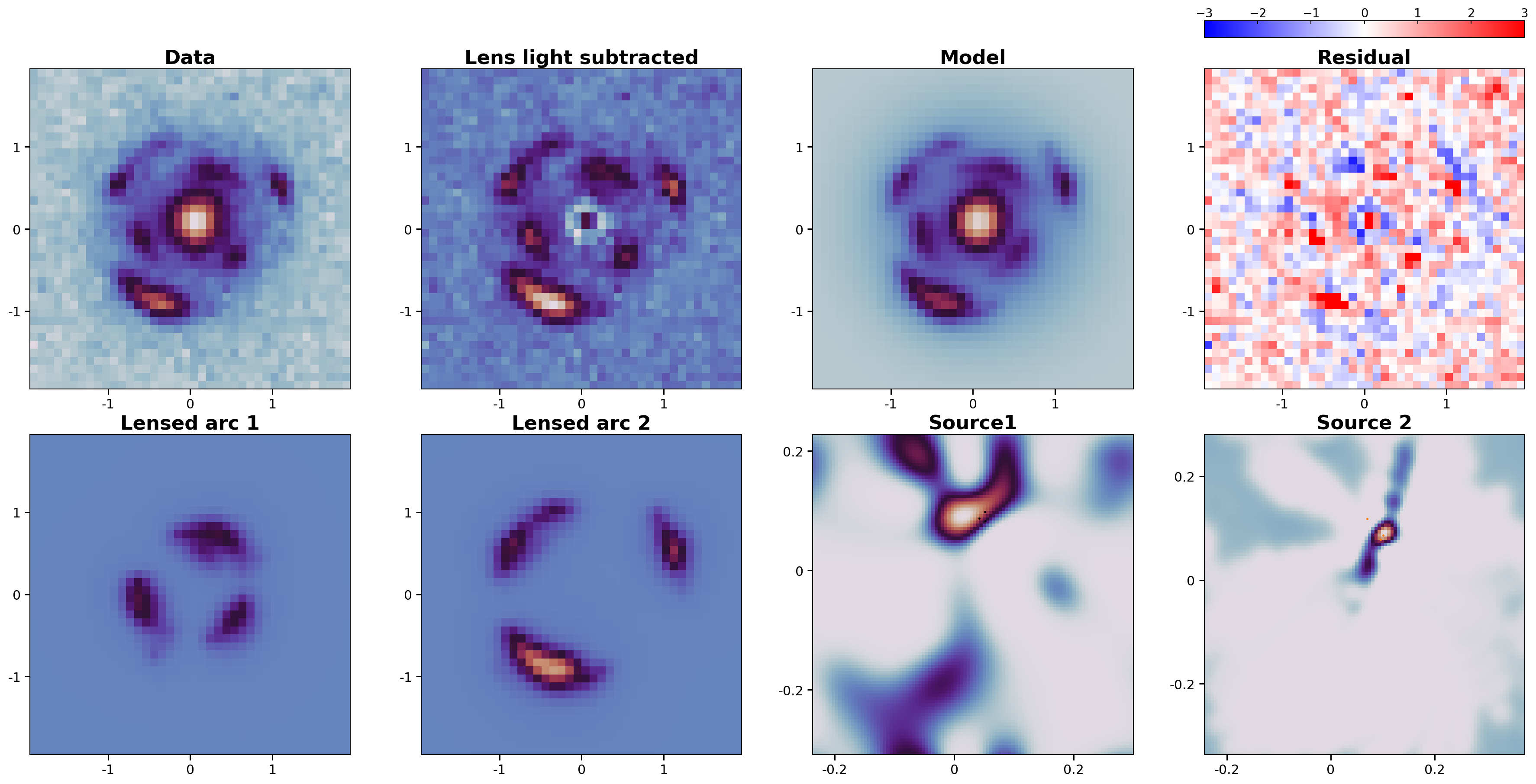}
    \caption{Alternative lens models for Silver candidate 47 (Euclid DR1\,J053233.630$-$371545.77), which can be interpreted either as a single-plane lens with multiple sources or as a DSPL. The \emph{top} panel shows the single-plane lens model, while the \emph{bottom} panel shows the DSPL lens model. Both models reproduce the observed configuration with relatively clean residuals, so the object remains in the Silver sample.}
    \label{fig:crackpotexample}
\end{figure*}

\subsection{Lens modelling strategy}

We used the open-source lens-modelling code \texttt{Herculens} \citep{Galan2022}.\footnote{\url{https://github.com/Herculens/herculens}} \texttt{Herculens} is based on the automatic differentiation and compilation features of \texttt{JAX} and can run on graphics processing units (GPUs).\footnote{\url{https://docs.jax.dev/en/latest/}} Here, we briefly describe the DSPL modelling used in this work and refer readers to \citet{Enzi2024} and \citet{Li2026jackpot} for more detailed accounts of this approach with \texttt{Herculens}. The purpose of our lens modelling is to assess whether these systems are plausible DSPLs, rather than to infer their parameters precisely. The lack of redshift information for the lenses and sources prevents us from constructing robust mass models or exploiting the full statistical power of Markov chain Monte Carlo (MCMC) methods.

The data we used for the modelling came from the \IE band because it has the highest pixel resolution (\ang{;;0.1}\,pixel$^{-1}$). We used the provided root mean square (RMS) maps and PSFs from the DR1 data products. The modelling was performed through the \texttt{NumPyro} implementation of SVI. We employed the \texttt{AdaBelief} optimiser \citep{Zhuang2020} with a low-rank multivariate normal distribution as the guiding probability distribution. This guiding distribution is not complex enough to accurately recover the full posterior distribution, but it is computationally much cheaper than MCMC methods because it can use the automatic-differentiation capabilities of \texttt{JAX} \citep{jax2018github}. The inference was carried out by optimising the guiding distribution to minimise the Kullback--Leibler divergence to the target posterior rather than sampling from it directly. This approach is computationally fast and can recover parameter point estimates close to those obtained from MCMC. However, the posterior standard deviations returned by SVI are substantially smaller than their MCMC counterparts and are therefore overconfident; we do not interpret them as reliable parameter uncertainties. Nevertheless, the resulting best-fitting models are sufficient to assess whether the observed image configurations are consistent with a DSPL interpretation.

Since the lensed images of source 1, which form the inner Einstein ring, and any images of source 2 close to the lens centre overlap with the lens light, we modelled the lens light simultaneously with the arcs. In our model, the lens light was modelled with the sum of several elliptical Gaussian luminosity profiles with flexible centres \citep{Shajib2019, He2024, Enzi2024}. The sources were first modelled with a single elliptical Gaussian luminosity profile to provide an approximate answer. Then, we used the parametrised lens model as the starting point and described the sources as fields that vary on regularly pixelated grids. The surface-brightness values on each pixel grid formed a vector $\boldsymbol{s}$. In Fourier space, its mode amplitudes were determined by a power spectrum \citep[e.g.,][]{Galan2024,RustigEtAl2024_LensCharm}:
\begin{equation}
\boldsymbol{s}=\mathcal{F}^{-1}\left[\sqrt{P_{\sfont{Mat}}(|\boldsymbol{k}|)}\odot\boldsymbol{\xi}\right]\;,
\end{equation}
where $\boldsymbol{k}$ is the wave vector of a pixel of the Fourier grid, $P_{\sfont{Mat}}(|\boldsymbol{k}|)$ is the isotropic Matérn power spectrum \citep{stein2012} evaluated at that pixel, $\odot$ denotes element-wise multiplication, $\boldsymbol{\xi}$ is standard Gaussian white noise whose elements are independent samples from $\mathcal{N}(0,1)$, and $\mathcal{F}^{-1}$ denotes the inverse Fourier transform. The default priors on the principal mass-model parameters are summarised in Table~\ref{tab:default_priors}. The upper bound on $\gamma_{\sfont{EPL}}$ excludes profiles steeper than $\gamma_{\sfont{EPL}}=2.8$, which do not give physical solutions for our systems. For some individual systems, we narrowed the prior range on $\gamma_{\sfont{EPL}}$ or $\theta_{\sfont{E},1}$ to obtain more stable lens models.

For each target, we ran between eight and ten SVI realisations, each with \num{20000} iterations. Each SVI realisation had a different random initialisation, and we selected the result with the lowest average loss value as our lens model.

\begin{figure}
    \centering
    \includegraphics[]{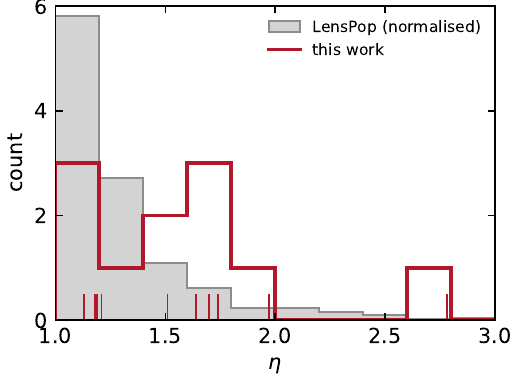}
    \caption{Distribution of $\eta$ inferred from the Gold DSPL lens models using the minimum-loss SVI realisation for each target, compared with the LensPop forecast of \ifdefined\PaperSixThesisMode\citet{Li2025b}\else\citet{Q1-SP054}\fi after removing systems with $\eta > 3$. The LensPop histogram is normalised to the total number of Gold systems, while the Gold histogram shows the observed counts from this work. The red ticks along the horizontal axis indicate the individual $\eta$ measurements for the Gold systems.}
    \label{fig:eta_dist}
\end{figure}

\section{Lens-model results}
\label{sec:result}

\begin{table*}
    \centering
    \setlength{\tabcolsep}{3pt}
    \renewcommand{\arraystretch}{1.25}
    \caption{DR1 Gold DSPL candidates with official source designations, sky coordinates, lens-model parameters from the selected best-SVI realisation for each target, \Euclid-band photometry, and available \Euclid-only photometric redshifts measured in a \ang{;;0.5} radius aperture centred on the \IE cutout. Candidates are numbered in order of increasing right ascension. The external-shear amplitude is $|\gamma_{\sfont{ext}}|=(\gamma_{\sfont{ext},1}^2+\gamma_{\sfont{ext},2}^2)^{1/2}$.}
    \label{tab:tier1_result}
    \begin{tabular}{clrrccccccccc}
    \hline
    No. & Official name & RA & Dec & $\theta_{\sfont{E},1}$ & $\gamma_{\sfont{EPL}}$ & $\eta$ & $\theta_{\sfont{E},\sfont{SIS}}$ & $|\gamma_{\sfont{ext}}|$ & \multicolumn{3}{c}{\IE magnitude\tablefootmark{a}} & \zph\tablefootmark{b}\\
    \cline{10-12}
     &  & $(^\circ)$ & $(^\circ)$ & $(\arcsec)$ & & & $(\arcsec)$ & & lens & s1 & s2 & lens \\
    \hline
    1 & Euclid DR1\,J035815.518$-$505648.90 & 59.5647 & $-50.9469$ & 0.48 & 2.15 & 2.78 & 0.66 & 0.003 & 21.14 & 22.35 & 23.13 & $0.84^{+0.05}_{-0.08}$ \\
    2 & Euclid DR1\,J040450.854$-$580447.18 & 61.2119 & $-58.0798$ & 1.03 & 2.57 & 1.97 & 0.05 & 0.012 & 18.60 & 22.46 & 23.37 & $0.40^{+0.04}_{-0.06}$ \\
    3 & Euclid DR1\,J040457.324$-$492221.54 & 61.2388 & $-49.3727$ & 1.16 & 2.21 & 1.18 & 0.21 & 0.082 & 19.72 & 22.05 & 23.67 & $0.54^{+0.01}_{-0.02}$ \\
    4 & Euclid DR1\,J044453.064$-$380845.60 & 71.2211 & $-38.1460$ & 1.58 & 1.80 & 1.64 & 0.19 & 0.244 & 21.30 & 22.53 & 23.61 & $0.84^{+0.05}_{-0.02}$ \\
    5 & Euclid DR1\,J045417.497$-$340950.95 & 73.5729 & $-34.1642$ & 0.79 & 1.63 & 1.51 & 0.16 & 0.100 & 19.80 & 21.89 & 23.79 & $0.52^{+0.05}_{-0.15}$ \\
    6 & Euclid DR1\,J060447.736$-$364412.03 & 91.1989 & $-36.7367$ & 1.81 & 2.19 & 1.51 & 0.28 & 0.033 & 19.39 & 23.70 & 23.49 & $0.46^{+0.09}_{-0.01}$ \\
    7 & Euclid DR1\,J115942.552+674249.68 & 179.9273 & 67.7138 & 1.43 & 1.36 & 1.13 & 0.04 & 0.105 & 16.73 & 22.29 & 21.73 & $0.36^{+0.01}_{-0.15}$ \\
    8 & Euclid DR1\,J125718.567+640613.04 & 194.3274 & 64.1036 & 1.00 & 1.88 & 1.70 & 0.28 & 0.017 & 17.12 & 20.43 & 20.79 & $0.10^{+0.02}_{-0.03}$ \\
    9 & Euclid DR1\,J172848.634+620851.61 & 262.2026 & 62.1477 & 0.71 & 1.65 & 1.74 & 0.16 & 0.034 & 20.79 & 22.88 & 23.27 & $0.91^{+0.01}_{-0.01}$ \\
    10 & Euclid DR1\,J181422.716+670816.26 & 273.5947 & 67.1379 & 0.65 & 1.49 & 1.19 & 0.05 & 0.152 & 19.68 & 23.24 & 23.23 & $0.59^{+0.02}_{-0.09}$ \\
    11 & Euclid DR1\,J181910.597+751312.57 & 274.7942 & 75.2202 & 0.93 & 2.19 & 1.21 & 0.40 & 0.120 & 20.05 & 22.52 & 21.62 & $0.50^{+0.01}_{-0.15}$ \\
    \hline
    \end{tabular}
    \tablefoot{
    \tablefoottext{a}{The lens light is fitted jointly with the mass model and the sources, so where it blends with the arcs the split of flux between deflector and sources is degenerate, and the SVI posterior underestimates this uncertainty.}
    \tablefoottext{b}{The photometric redshifts are indicative only. Their quoted uncertainties are severely underestimated: they propagate only the background noise in the aperture and include no allowance for the PSF mismatch between \IE and the NISP bands, for contamination of the aperture by the lensed arcs, or for template mismatch.}
    }
\end{table*}

\begin{table}[!t]
    \centering
    \caption{Default priors for the principal DSPL mass-model parameters. Here $\mathcal{U}(a,b)$ denotes a uniform distribution between $a$ and $b$, and $\mathrm{TN}_{[a,b]}(\mu,\sigma)$ denotes a normal distribution with mean $\mu$ and standard deviation $\sigma$, truncated to $[a,b]$.}
    \label{tab:default_priors}
    \begin{tabular}{ll}
    \hline\hline
    Parameter & Prior \\
    \hline
    $\theta_{\sfont{E},1}$ & $\mathcal{U}(0,3\,\mathrm{arcsec})$ \\
    $e_1$, $e_2$ & $\mathrm{TN}_{[-0.2,0.2]}(0,0.25)$ \\
    $\eta$ & $\mathcal{U}(1,3)$ \\
    $\theta_{\sfont{E},\sfont{SIS}}$ & $\mathcal{U}(0,1\,\mathrm{arcsec})$ \\
    $\gamma_{\sfont{EPL}}$ & $\mathcal{U}(1.2,2.8)$ \\
    $\gamma_{\sfont{ext},1}$, $\gamma_{\sfont{ext},2}$ & $\mathcal{U}(-0.2,0.2)$ \\
    \hline
    \end{tabular}
\end{table}

We classified \sampleprovisional{84} DSPL candidates and divided them into two classes, Gold and Silver. The Gold sample consists of \sampleprovisional{11} systems in which the multiple images produced by each of the two sources are clearly identifiable and can be well reproduced by our lens models. Colour images of the Gold DSPL candidates are shown in Fig.~\ref{fig:Tier1image}, an example lens model is shown in Fig.~\ref{fig:tier1example}, and the complete set of Gold lens-model reconstructions is shown in Appendix~\ref{app:tier1_lens_models}. Integral-field-unit (IFU) spectroscopy of these targets would measure the redshifts of all components and is the most efficient route to confirming them as DSPLs.

The Gold lens modelling results are presented in Table~\ref{tab:tier1_result}. The quoted \IE magnitudes are measured from the \IE-band images. For the deflector, we integrated the total flux of the best-fitting multi-Gaussian expansion (MGE) lens-light model and converted it to an AB magnitude using the \IE-band zeropoint. For the two sources, we first subtracted the MGE lens-light model from the \IE image and then summed the residual flux inside the corresponding arc masks used in the lens modelling. The lens photometric redshifts are \Euclid-only template-fitting estimates computed with \texttt{Phosphoros}, the photometric-redshift code used by the \Euclid photometric-redshift (PHZ) processing function \citep{Desprez-EP10,Q1-TP005}. We used aperture photometry in a \ang{;;0.5} radius aperture centred on the \IE cutout, with the available \IE, \YE, \JE, and \HE bands. The quoted redshift uncertainties are the 16th--84th percentile interval of the one-dimensional redshift probability density function (PDF). Photometric redshifts for the full DR1 strong-lens sample, derived with a more complete treatment than the simple template fit used here, are presented by Euclid Collaboration: Nersesian et al. (in prep.). We did not attempt photometric redshifts for the sources. The arcs are resolved from the deflector only in \IE; in the NISP bands, whose PSF is about three times broader, they blend with the much brighter lens light. Reliable arc photometry would therefore require a joint multi-band model of the lens light and the sources, which is beyond the scope of this work, and \IE photometry alone cannot constrain the source redshifts.

Some systems have a large value of $\theta_{\sfont{E},\sfont{SIS}}$ ($\theta_{\sfont{E},\sfont{SIS}}>\ang{;;0.5}$), which describes the lensing strength of source 1 on source 2. There are two reasons why this can occur. First, the values in Table~\ref{tab:tier1_result} are the SVI medians from the selected run. SVI gives posterior standard deviations that are too small, so these values should not be treated as precise measurements. In some systems, the light rays from source 2 pass far from source 1, so the data do not constrain the mass of source 1 well enough. The result can then depend on the SVI guide and random initialisation, and a much smaller value of $\theta_{\sfont{E},\sfont{SIS}}$ may also fit the data. Second, source 1 may have a genuinely large lensing strength. Our search tends to select systems with two rings of very different radii. A large outer ring can therefore favour systems in which source 1 makes a strong contribution to the lensing of source 2.

We classify Silver DSPL candidates based on several forms of ambiguity. Nine additional Silver candidates were selected by visual inspection and have not yet been lens-modelled. Some systems contain one complete Einstein ring and an additional lensed arc with no counter image, while others contain two lensed arcs but no clear multiple-image configuration. Robust lens modelling is more challenging in these cases because the image-plane constraints are incomplete. We also retain systems for which a single-plane lens model with multiple sources can reproduce the observed configuration, because the current data do not establish that the components lie at distinct source redshifts. Euclid DR1\,J023930.713$-$515748.30 is included in Silver because its best-fitting $\eta=1.01$ is very close to the single-source-plane limit. Three systems in our sample are also reported in Q1 \ifdefined\PaperSixThesisMode\citep{Li2025b}\else\citep{Q1-SP054}\fi, and Silver candidate 28, DES~J0408$-$5354, was previously discovered in DES by \citet{2017ApJ...838L..15L}; the remaining \sampleprovisional{80} candidates are newly identified here. DES~J0408$-$5354 is a DSPL very close to the single-plane limit: MUSE spectroscopy places its second source at $z=2.228$, close to the quasar at $z=2.375$ \citep{Shajib2020STRIDES}, giving $\eta\simeq1.02$. The Silver colour images are shown together in Fig.~\ref{fig:Tier2image}, examples of the incomplete-image and single-plane-compatible cases are shown in Figs.~\ref{fig:Tier2example} and \ref{fig:crackpotexample}, respectively, and their coordinates are presented in Table~\ref{tab:tier2_ra_dec}. Several lens candidates also include an arc that is difficult to model and difficult to explain as an astrophysical contaminant.

\section{Discussion}
\label{sec:discussion}

Several objects in our sample have ambiguous morphologies, and their physical nature remains difficult to determine from the current \Euclid imaging and lens modelling alone.
\begin{itemize}
\item Silver candidate 70, Euclid DR1\,J184618.330+723826.77, is a particularly interesting system. It consists of an elliptical galaxy, a nearby satellite galaxy, and several blue arc-like features. These features have morphologies that are typical of gravitationally lensed arcs. However, the system contains nine apparent images in total, and their spatial distribution is difficult to explain with a standard strong-lensing configuration. Their colours are also very similar, making it difficult to assign the different images to distinct source-plane positions. We therefore keep this object in the Silver sample.

\item Silver candidate 32, Euclid DR1\,J041020.748$-$441418.62, is another difficult lens system to model. Its inner ring can be well reproduced by a gravitational lens model, but the outer blue structure is much harder to interpret. This outer structure does not resemble a conventional spiral arm, nor does it look like a nearby galaxy undergoing tidal interaction. Because its nature remains unclear, we classify it as a Silver candidate.

\item Silver candidate 10, Euclid DR1\,J022455.052$-$601235.97, is a promising disc-galaxy DSPL candidate. The image shows multiple ring-like structures, including symmetric arc-like features above and below the lens, as well as additional arc-like structures on the left and right sides of the disc core. These features are characteristic of strong gravitational lensing. However, the lens light contains a prominent dust lane that overlaps with the arcs, and the outer arc is very faint. As a result, our DSPL models for this system are not stable: different SVI chains converge to different configurations. Given this uncertainty, we classify this object as a Silver lens candidate. Deeper imaging at higher angular resolution, for example with HST or JWST, would be valuable for this system.

\item Silver candidate 47, Euclid DR1\,J053233.630$-$371545.77, features a central galaxy with six different images. We performed both DSPL and single-plane lens models. The modelling results are shown in Fig.~\ref{fig:crackpotexample}. Both lens models can reproduce the configuration of the data with relatively clean residuals. The single-plane model has an Einstein radius of \ang{;;0.70}. The DSPL model has an Einstein radius of the first ring of \ang{;;0.70} and $\eta=1.52$.
\end{itemize}

\section{Limitations and comparison to expectations}
\label{sec:forecast}

Our Q1 study \ifdefined\PaperSixThesisMode\citep{Li2025b}\else\citep{Q1-SP054}\fi{} used the LensPop code \citep{Collett:2015roa} to forecast that \Euclid will discover approximately 1700 galaxy-scale DSPLs over the full survey area. Scaling this prediction to the 1900\,deg$^2$ covered by DR1 gives an expectation of about 200 DSPLs. Alternatively, extrapolating from the three robust DSPL candidates identified in Q1 leads to an expectation of about 90 DSPL systems in DR1. Both numbers are substantially higher than the number of Gold DSPL candidates found in this work. Even if all of the Silver DSPL candidates are real DSPLs, there remains a large shortfall relative to the forecasts. We attribute the difference to two effects: the limitations of our search strategy and the definition of a DSPL. Figure~\ref{fig:eta_dist} compares the distribution of $\eta$ in our Gold lens sample with this LensPop forecast. The forecast sample is truncated at $\eta=3$, the upper bound of our $\eta$ prior (Table~\ref{tab:default_priors}). Only 13 of the 1729 forecast systems (0.8\%) have $\eta>3$. In these systems source 1 lies just behind the deflector (median $z_{\mathrm{s1}}-z_{\mathrm{d}}=0.12$), so the inner Einstein radius is very small (median \ang{;;0.25}), close to the VIS PSF and embedded in the lens light, and such systems are unlikely to be recognised in \Euclid imaging. The median $\eta$ of the Gold sample is 1.51, compared with 1.18 for the LensPop forecast, and 8 of the 11 Gold systems (73\%) have $\eta>1.2$, whereas only 47\% of the forecast systems do. Because our candidates were selected by visual inspection, our selection naturally favours large $\eta$: when $\eta$ is close to unity, the two Einstein rings nearly coincide and blend together at the \Euclid resolution, as in Silver candidates 9 and 60 (Fig.~\ref{fig:Tier2image}). Such systems are either classified as Silver or missed by our visual inspection.

Our search is limited in that we did not systematically search for DSPLs: we did not train our machine-learning classifiers to identify DSPLs. Instead, we visually inspected the best-ranked strong-lens candidates identified by the DR1 strong-lens finding networks (Euclid Collaboration: Vincken et al., in prep.; Euclid Collaboration: Fogliardi et al., in prep.) and prioritised through the ECLIPSE discovery pipeline (Euclid Collaboration: Holloway et al., in prep.). We do not know how many DSPLs are ranked below these visual-inspection thresholds. Additionally, we primarily searched for DSPLs in the same colour-composite images, so faint second sources are likely to have been missed. Whilst some DSPL candidates are identified in lens-light-subtracted residuals from the mass models of Euclid Collaboration: Nightingale et al. (in prep.), we did not systematically calibrate this. Another family of missing DSPL should be systems with source that lacks VIS image as they are at high redshift or dusty.

The DSPL definition adopted in LensPop is based on the length of the arc: a source galaxy is classified as lensed if the arc length is larger than \ang{;;0.3}. However, this does not necessarily imply the formation of a bright and obvious counter image. Several of our Silver DSPL candidates fall into this category: the second source is likely substantially magnified, but it does not form a clear multiple-image configuration. For these systems, it is difficult to use the second source plane to provide additional constraints on the lens model. As discussed by \ifdefined\PaperSixThesisMode\citet{Tian2026}\else\citet{Li2026jackpot}\fi, the key advantage of a DSPL is that the second Einstein ring provides an additional enclosed-mass measurement, which improves constraints on the radial mass profile. If the second source does not form multiple images, however, constraints on the deflection field are much weaker and almost all information about the lens mass profile comes from the first complete Einstein ring. Furthermore, because the first source plane may itself contain non-negligible mass, the intrinsic morphology of the second source cannot be robustly constrained without a clear multiple-image configuration.

\section{Conclusions}
\label{sec:conclusion}
We present \sampleprovisional{84} galaxy-scale DSPL candidates in DR1, consisting of \sampleprovisional{11} Gold and \sampleprovisional{73} Silver systems; \sampleprovisional{80} of these candidates are newly identified. This substantially enlarges the published sample of galaxy-scale DSPL candidates. The detailed lens modelling shows that the modelled candidates can be explained by a DSPL configuration, although some Silver systems are also compatible with a single-plane interpretation. We performed lens modelling and attempted to derive $\eta$ for the candidates with available models; nine additional Silver candidates remain to be modelled. For the Silver candidates, higher-resolution imaging and spectroscopic redshifts are needed for a definitive conclusion about their nature.

The shortfall relative to the forecasts shows that many DSPLs remain undetected in DR1, either because they were assigned low rankings by the single-source-plane lens-finding classifiers or because they were missed during visual inspection. It is clear that piggybacking on the efforts to find galaxy--galaxy lenses in \Euclid can never provide a high-completeness DSPL search: if \Euclid is to reach its full DSPL discovery potential, then a dedicated search is needed, including training DSPL-specific machine-learning finders, automated lens modelling of the candidates, and teaching citizen scientists to look out for them.

%
%
The \sampleprovisional{84} candidates form a valuable sample from which to select individual high-value systems for follow-up. In each such system, the additional radial constraints from a second source plane improve the separation of the stellar and dark matter components and allow assumptions about the stellar initial mass function (IMF) to be tested. For the Jackpot DSPL, \citet{Li2026jackpot} constrain the IMF mismatch parameter from lensing alone to $\alpha_{\sfont{SPS}}=0.91^{+0.17}_{-0.15}$, sufficient to distinguish its Salpeter-like IMF from a Chabrier IMF at approximately $3\sigma$. A sample of around 50 well-constrained DSPLs spanning a broad redshift range could therefore test whether the IMF normalisation is universal and whether it evolves with redshift. For comparison, Project Dinos II combines strong lensing and stellar dynamics for 56 lenses and constrains the redshift-dependence parameter of the dark matter inner slope to $\alpha_{\gamma_{\sfont{in}}}=-0.44^{+0.14}_{-0.15}$ \citep{Sheu2025}. A similarly sized DSPL sample could test the evolution of dark matter density profiles using the additional lensing constraints from the second source plane. DSPLs can also probe the mass associated with source 1 through its lensing effect on source 2, providing information about lower-mass galaxies at higher redshift.

\ifdefined\PaperSixThesisMode
\else
\begin{acknowledgements}
This project and WJRE have received funding from the European Research Council (ERC) under the European Union's Horizon 2020 research and innovation programme (LensEra: grant agreement No 945536). TEC is funded by the Royal Society through a University Research Fellowship. WJRE acknowledges the support by the Italian Ministry of University and Research (grant FIS2023-01611, CUP C53C25000300001).

NL is supported through a graduate studentship from the UKRI STFC and the University of Portsmouth, and acknowledges travel support from STFC for UK participation in the Rubin LSST through grant ST/X001334/1. PH is funded by the Royal Society (RF\textbackslash{}ERE\textbackslash{}210033). JWN is supported by an STFC/UKRI Ernest Rutherford Fellowship, Project Reference: ST/X003086/1. NL, PH, and JWN acknowledge travel support provided by Euclid:UK through STFC grant APP91126.

AG acknowledges support by the SNSF through PostDoc Mobility Return grant P5R5-2\_235360. AS acknowledges support from the National Key R\&D Program of China (No. 2023YFA1607800, 2023YFA1607802), and from the Yangyang Development Fund.

Numerical computations were done on the Sciama High Performance Compute (HPC) cluster which is supported by the ICG, SEPNet and the University of Portsmouth.

\AckDRone
\AckEC
\end{acknowledgements}

%
%

\bibliography{Euclid, DR1, mycitation} 
%
%

%
\fi


\ifdefined\PaperSixThesisMode
\else
\begin{appendix}
  \onecolumn
\fi

\section{Gold lens-model reconstructions}
\label{app:tier1_lens_models}

\begin{figure}[!ht]
    \centering
    \labeledsixpanelmodelimage{1: Euclid DR1\,J035815.518$-$505648.90}{1}{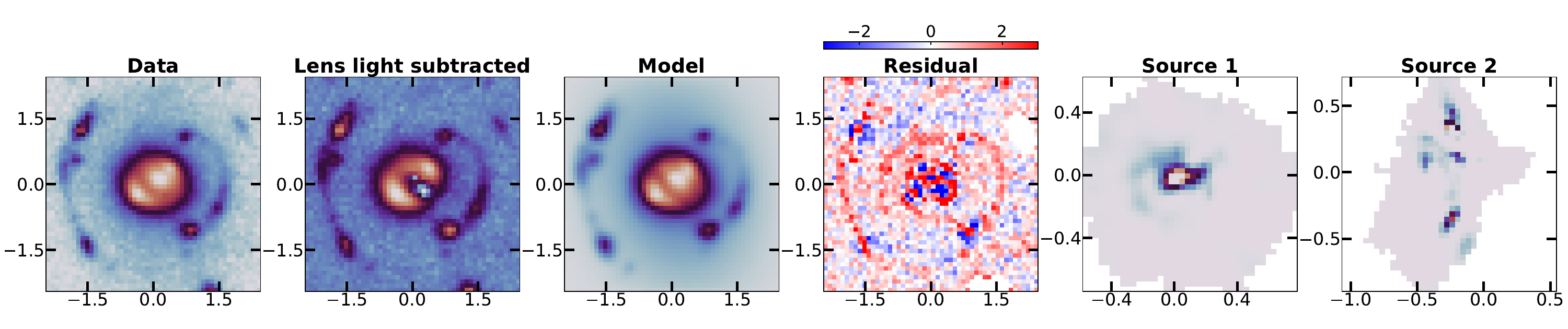}
    \labeledsixpanelmodelimage{2: Euclid DR1\,J040450.854$-$580447.18}{1}{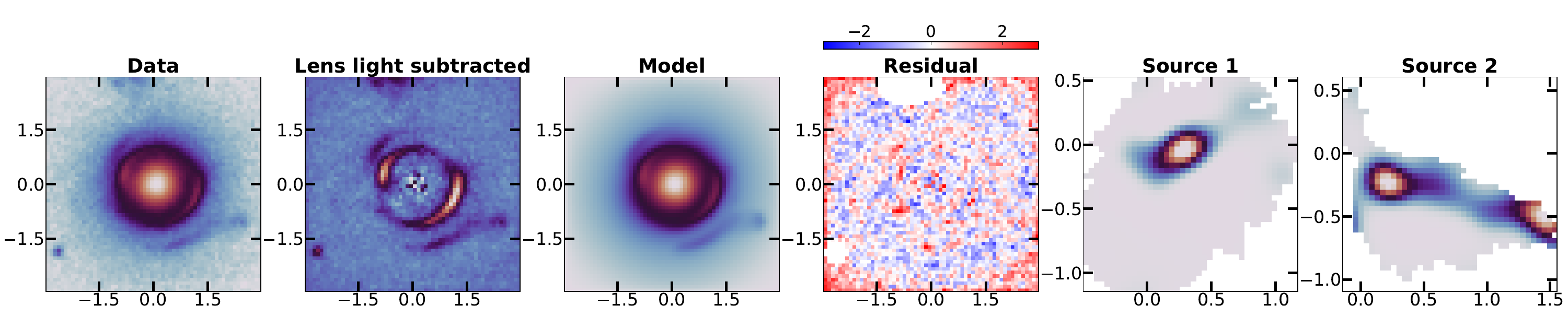}
    \labeledsixpanelmodelimage{3: Euclid DR1\,J040457.324$-$492221.54}{1}{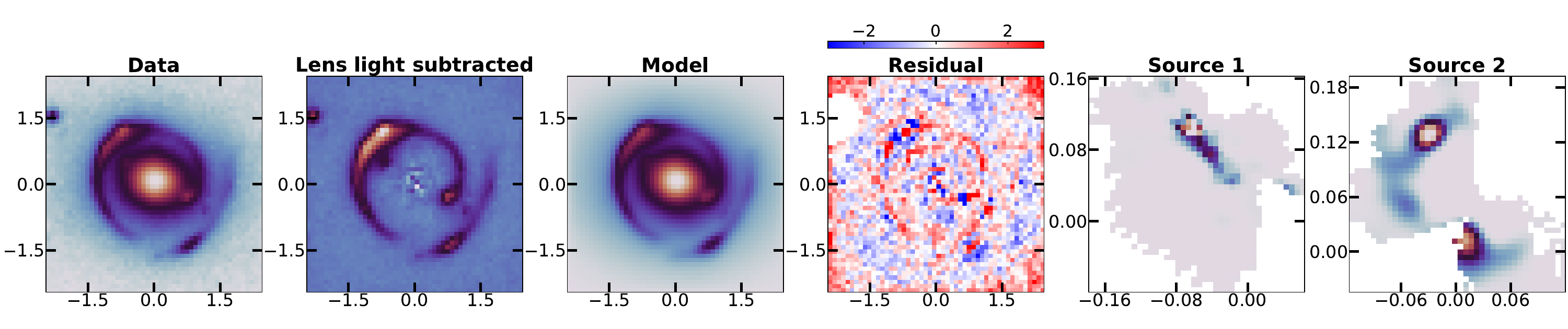}
    \labeledsixpanelmodelimage{4: Euclid DR1\,J044453.064$-$380845.60}{1}{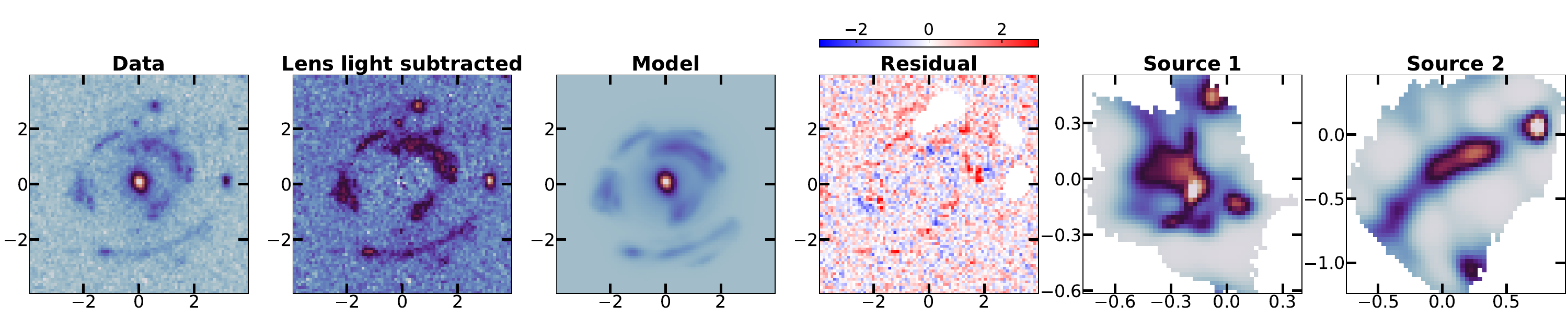}
    \labeledsixpanelmodelimage{5: Euclid DR1\,J045417.497$-$340950.95}{1}{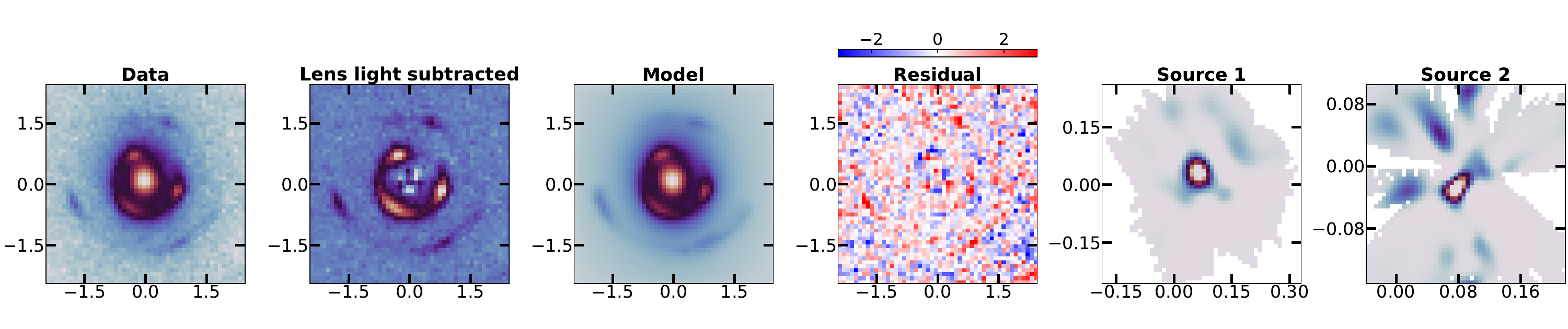}
    \labeledsixpanelmodelimage{6: Euclid DR1\,J060447.736$-$364412.03}{1}{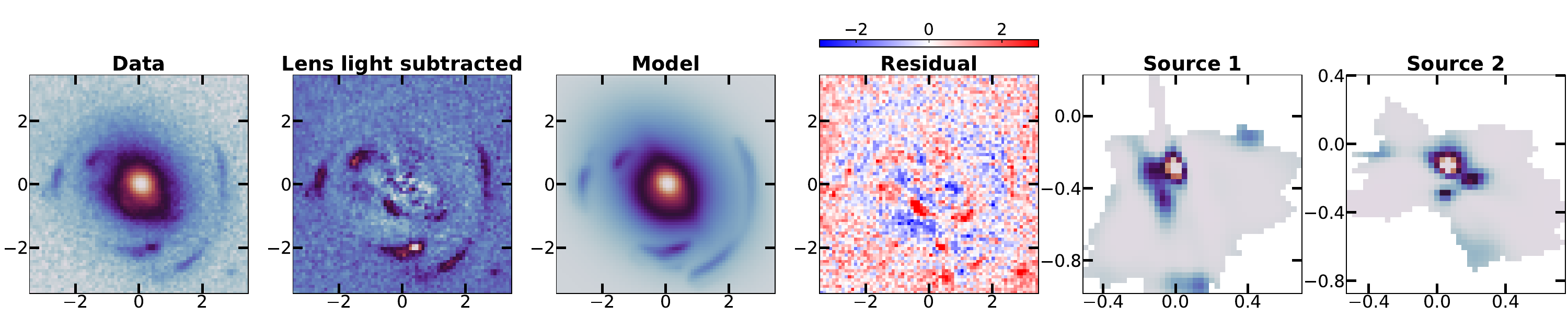}
    \caption{Lens-model reconstructions for Gold candidates 1--6. From \emph{left} to \emph{right}, the panels show the observed data, lens-light-subtracted data, reconstructed model, normalised residuals, and the pixelated reconstructions of source 1 and source 2.}
    \label{fig:tier1_lens_models_01_05}
\end{figure}

\begin{figure}[p]
    \centering
    \labeledsixpanelmodelimage{7: Euclid DR1\,J115942.552+674249.68}{1}{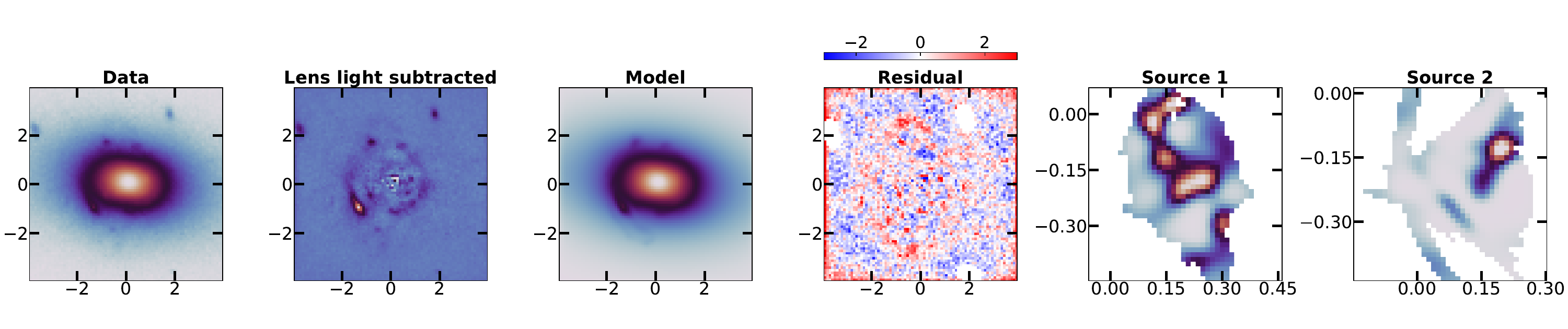}
    \labeledsixpanelmodelimage{8: Euclid DR1\,J125718.567+640613.04}{1}{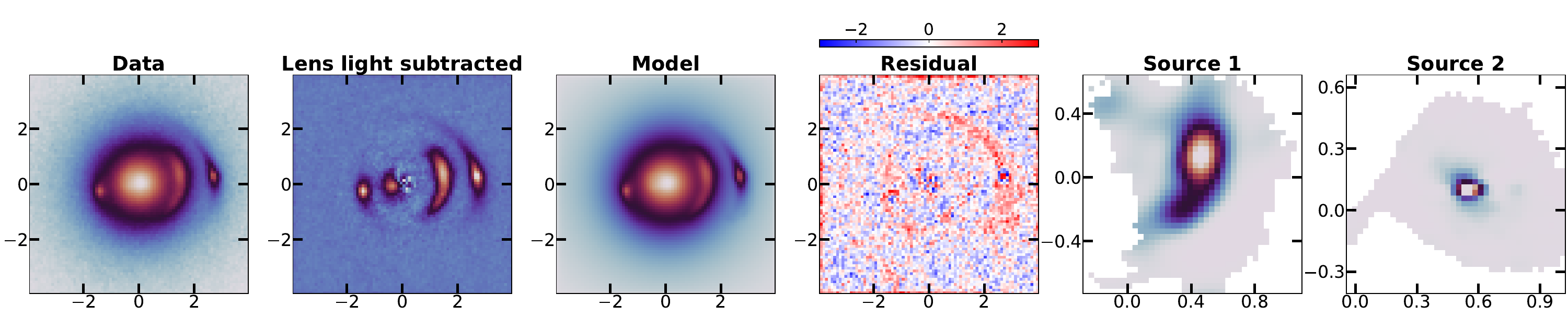}
    \labeledsixpanelmodelimage{9: Euclid DR1\,J172848.634+620851.61}{1}{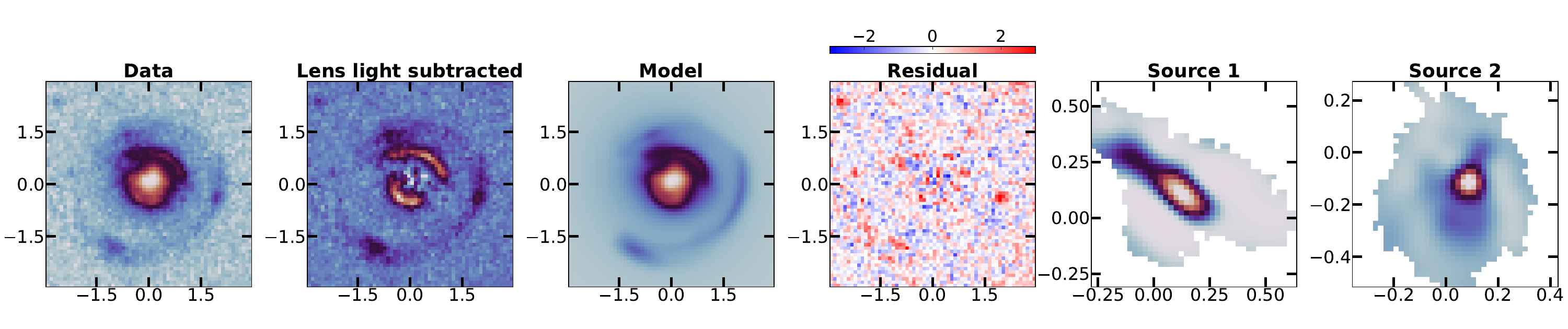}
    \labeledsixpanelmodelimage{10: Euclid DR1\,J181422.716+670816.26}{1}{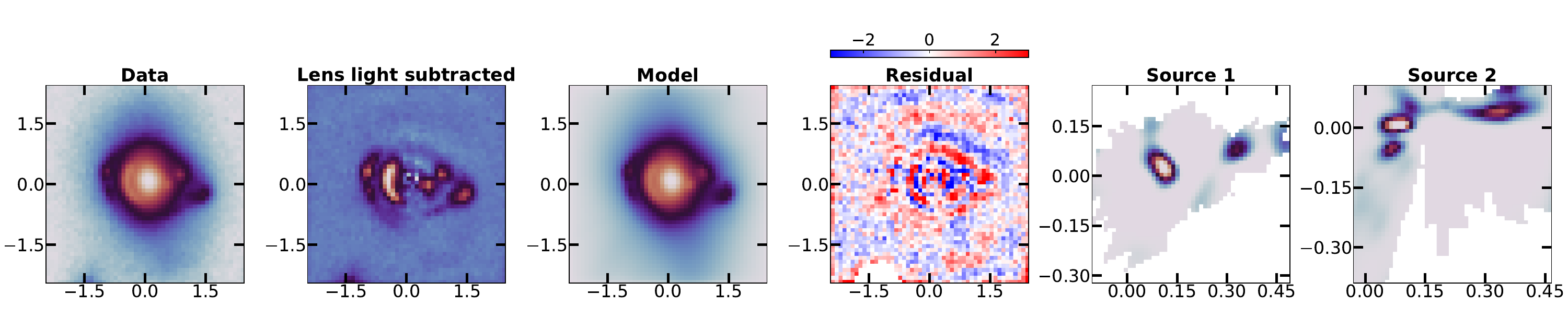}
    \labeledsixpanelmodelimage{11: Euclid DR1\,J181910.597+751312.57}{1}{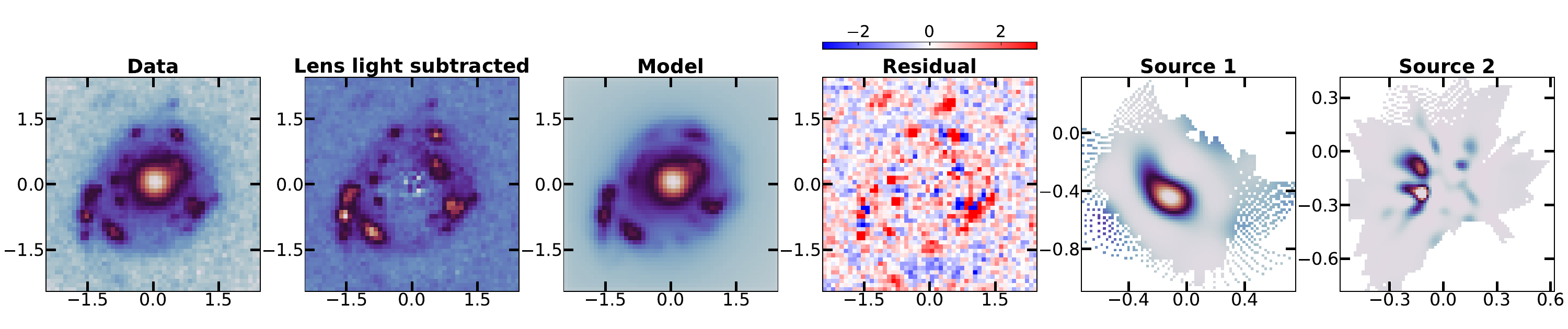}
    \caption{Lens-model reconstructions for Gold candidates 7--11. The panels follow the layout described in Fig.~\ref{fig:tier1_lens_models_01_05}.}
    \label{fig:tier1_lens_models_06_11}
\end{figure}
\clearpage

\begin{table}[!t]
    \centering
    \footnotesize
    \setlength{\tabcolsep}{3pt}
    \renewcommand{\arraystretch}{1.15}
    \caption{DR1 Silver DSPL candidates with official source designations and sky coordinates. Candidates are numbered in order of increasing right ascension, and the numbering follows the labels printed in Fig.~\ref{fig:Tier2image}.}
    \label{tab:tier2_ra_dec}
    \begin{tabular}{rlrr@{\hspace{14pt}}rlrr}
    \hline
    \multicolumn{4}{c}{Silver candidates 1--37} & \multicolumn{4}{c}{Silver candidates 38--73} \\
    \hline
    No. & Official name & RA & Dec & No. & Official name & RA & Dec \\
     &  & $(^\circ)$ & $(^\circ)$ &  &  & $(^\circ)$ & $(^\circ)$ \\
    \hline
    1 & Euclid DR1\,J004938.260$-$533600.41 & 12.4094 & $-53.6001$ & 38 & Euclid DR1\,J042829.614$-$352625.48 & 67.1234 & $-35.4404$ \\
    2 & Euclid DR1\,J005136.896$-$541119.61 & 12.9037 & $-54.1888$ & 39 & Euclid DR1\,J043010.921$-$241017.92 & 67.5455 & $-24.1716$ \\
    3 & Euclid DR1\,J015016.504$-$475016.16 & 27.5688 & $-47.8378$ & 40 & Euclid DR1\,J043302.910$-$495131.42 & 68.2621 & $-49.8587$ \\
    4 & Euclid DR1\,J015306.152$-$432312.45 & 28.2756 & $-43.3868$ & 41 & Euclid DR1\,J044353.266$-$225204.10 & 70.9719 & $-22.8678$ \\
    5 & Euclid DR1\,J015716.307$-$415630.25 & 29.3179 & $-41.9417$ & 42 & Euclid DR1\,J044448.827$-$381852.90 & 71.2034 & $-38.3147$ \\
    6 & Euclid DR1\,J015749.436$-$453740.42 & 29.4560 & $-45.6279$ & 43 & Euclid DR1\,J050145.976$-$411800.54 & 75.4416 & $-41.3002$ \\
    7 & Euclid DR1\,J020203.314$-$561048.81 & 30.5138 & $-56.1802$ & 44 & Euclid DR1\,J050312.533$-$301125.81 & 75.8022 & $-30.1905$ \\
    8 & Euclid DR1\,J020636.832$-$475440.27 & 31.6535 & $-47.9112$ & 45 & Euclid DR1\,J050456.363$-$340032.88 & 76.2348 & $-34.0091$ \\
    9 & Euclid DR1\,J020712.749$-$314638.30 & 31.8031 & $-31.7773$ & 46 & Euclid DR1\,J051447.016$-$405357.27 & 78.6959 & $-40.8992$ \\
    10 & Euclid DR1\,J022455.052$-$601235.97 & 36.2294 & $-60.2100$ & 47 & Euclid DR1\,J053233.630$-$371545.77 & 83.1401 & $-37.2627$ \\
    11 & Euclid DR1\,J022748.280$-$385955.46 & 36.9512 & $-38.9987$ & 48 & Euclid DR1\,J055925.749$-$520844.09 & 89.8573 & $-52.1456$ \\
    12 & Euclid DR1\,J023930.713$-$515748.30 & 39.8780 & $-51.9634$ & 49 & Euclid DR1\,J062002.124$-$491417.05 & 95.0089 & $-49.2381$ \\
    13 & Euclid DR1\,J024734.522$-$515450.94 & 41.8938 & $-51.9141$ & 50 & Euclid DR1\,J100050.272+013251.75 & 150.2095 & 1.5477 \\
    14 & Euclid DR1\,J025109.558$-$612310.57 & 42.7898 & $-61.3863$ & 51 & Euclid DR1\,J103048.472+755521.17 & 157.7020 & 75.9225 \\
    15 & Euclid DR1\,J030425.787$-$473743.72 & 46.1074 & $-47.6288$ & 52 & Euclid DR1\,J103633.039+683649.44 & 159.1377 & 68.6137 \\
    16 & Euclid DR1\,J030718.758$-$422841.88 & 46.8282 & $-42.4783$ & 53 & Euclid DR1\,J104003.271+742251.11 & 160.0136 & 74.3809 \\
    17 & Euclid DR1\,J031158.590$-$463445.16 & 47.9941 & $-46.5792$ & 54 & Euclid DR1\,J105143.454+724215.93 & 162.9311 & 72.7044 \\
    18 & Euclid DR1\,J031543.262$-$622035.56 & 48.9303 & $-62.3432$ & 55 & Euclid DR1\,J122728.143+702548.39 & 186.8673 & 70.4301 \\
    19 & Euclid DR1\,J032408.363$-$452614.20 & 51.0348 & $-45.4373$ & 56 & Euclid DR1\,J124443.089+644001.66 & 191.1795 & 64.6671 \\
    20 & Euclid DR1\,J032732.056$-$364503.45 & 51.8836 & $-36.7510$ & 57 & Euclid DR1\,J124942.621+661313.49 & 192.4276 & 66.2204 \\
    21 & Euclid DR1\,J032801.253$-$402957.18 & 52.0052 & $-40.4992$ & 58 & Euclid DR1\,J142643.320+441110.32 & 216.6805 & 44.1862 \\
    22 & Euclid DR1\,J033258.327$-$571551.80 & 53.2430 & $-57.2644$ & 59 & Euclid DR1\,J142652.199+433113.80 & 216.7175 & 43.5205 \\
    23 & Euclid DR1\,J033611.349$-$574906.11 & 54.0473 & $-57.8184$ & 60 & Euclid DR1\,J150758.546+724738.33 & 226.9939 & 72.7940 \\
    24 & Euclid DR1\,J033709.402$-$275811.26 & 54.2892 & $-27.9698$ & 61 & Euclid DR1\,J151918.638+721127.70 & 229.8277 & 72.1910 \\
    25 & Euclid DR1\,J034234.580$-$412024.11 & 55.6441 & $-41.3400$ & 62 & Euclid DR1\,J171314.881+602207.86 & 258.3120 & 60.3688 \\
    26 & Euclid DR1\,J034704.604$-$451603.69 & 56.7692 & $-45.2677$ & 63 & Euclid DR1\,J173609.281+605546.61 & 264.0387 & 60.9296 \\
    27 & Euclid DR1\,J040105.630$-$354724.50 & 60.2734 & $-35.7901$ & 64 & Euclid DR1\,J173839.271+610400.88 & 264.6636 & 61.0669 \\
    28 & Euclid DR1\,J040821.624$-$535359.10 & 62.0901 & $-53.8998$ & 65 & Euclid DR1\,J174745.539+640250.16 & 266.9397 & 64.0473 \\
    29 & Euclid DR1\,J040920.805$-$470401.02 & 62.3367 & $-47.0670$ & 66 & Euclid DR1\,J175229.144+610937.80 & 268.1214 & 61.1605 \\
    30 & Euclid DR1\,J041000.944$-$510607.51 & 62.5039 & $-51.1021$ & 67 & Euclid DR1\,J175446.694+713921.20 & 268.6946 & 71.6559 \\
    31 & Euclid DR1\,J041013.488$-$552227.26 & 62.5562 & $-55.3742$ & 68 & Euclid DR1\,J175746.354+651109.29 & 269.4431 & 65.1859 \\
    32 & Euclid DR1\,J041020.748$-$441418.62 & 62.5864 & $-44.2385$ & 69 & Euclid DR1\,J181935.805+570832.64 & 274.8992 & 57.1424 \\
    33 & Euclid DR1\,J041739.371$-$301716.04 & 64.4140 & $-30.2878$ & 70 & Euclid DR1\,J184618.330+723826.77 & 281.5764 & 72.6408 \\
    34 & Euclid DR1\,J041906.519$-$295937.60 & 64.7772 & $-29.9938$ & 71 & Euclid DR1\,J233054.472$-$330559.85 & 352.7270 & $-33.1000$ \\
    35 & Euclid DR1\,J042037.009$-$341447.85 & 65.1542 & $-34.2466$ & 72 & Euclid DR1\,J234338.579$-$335641.58 & 355.9107 & $-33.9449$ \\
    36 & Euclid DR1\,J042532.089$-$482052.54 & 66.3837 & $-48.3479$ & 73 & Euclid DR1\,J235319.516$-$602119.86 & 358.3313 & $-60.3555$ \\
    37 & Euclid DR1\,J042724.367$-$233942.21 & 66.8515 & $-23.6617$ &  &  &  &  \\
    \hline
    \end{tabular}
\end{table}

\ifdefined\PaperSixThesisMode
\else
\end{appendix}

\label{LastPage}
\end{document}
\fi